\documentclass[aps,prb,twocolumn,superscriptaddress]{revtex4-1} 
\usepackage{array,mathtools,amssymb,booktabs} 
\usepackage{graphicx}
\usepackage[pdftex,colorlinks,citecolor=blue,linkcolor=blue,urlcolor=blue]{hyperref} 
\usepackage{amsfonts}
\usepackage{subfigure}
\usepackage{bm}
\usepackage{multirow}
\usepackage{stmaryrd}
\usepackage{latexsym}

\newcommand{\be}{\begin{equation}}
\newcommand{\ee}{\end{equation}}
\newcommand{\bea}{\begin{eqnarray}}
\newcommand{\eea}{\end{eqnarray}}
\newcommand{\barr}{\begin{array}}
\newcommand{\earr}{\end{array}}
\newcommand{\figref}[1]{Fig.\,\ref{#1}}
\long\def\begincomment#1\endcomment{} 

\newcommand{\diag}{\mathop{\mathrm{diag}}}
\newcommand{\rank}{\mathop{\mathrm{rk}}}

\newcommand{\GSD}{\mathop{\mathrm{GSD}}}

\newcommand{\even}{\mathop{\mathrm{even}}}
\newcommand{\odd}{\mathop{\mathrm{odd}}}
\newcommand{\ti}{\mathrm{i}}

\usepackage{comment} 

\newcolumntype{C}{>{$}c<{$}}

\begin{document}

\title{Boundary Degeneracy of Topological Order} %
\author{Juven C. Wang} \email{juven@mit.edu} 
\affiliation{Department of Physics, Massachusetts Institute of Technology, Cambridge, MA 02139, USA}
\affiliation{Perimeter Institute for Theoretical Physics, Waterloo, ON, N2L 2Y5, Canada}
\author{Xiao-Gang Wen} \email{wen@dao.mit.edu} 
\affiliation{Perimeter Institute for Theoretical Physics, Waterloo, ON, N2L 2Y5, Canada}
\affiliation{Department of Physics, Massachusetts Institute of Technology, Cambridge, MA 02139, USA}
\affiliation{Institute for Advanced Study, Tsinghua University, Beijing,
100084, P. R. China}



\begin{abstract}

We introduce the 
 concept of boundary  
degeneracy 
of 
topologically ordered states on a compact orientable spatial manifold with boundaries, 
and emphasize that  
the boundary degeneracy provides richer information than the bulk degeneracy.
Beyond the bulk-edge correspondence,
we find the ground state degeneracy of the fully gapped  
edge modes depends on boundary gapping conditions. 
By associating different types of boundary gapping conditions
as different ways of 
particle or quasiparticle condensations on the boundary,
we develop an analytic
theory of gapped boundaries. 
By Chern-Simons theory, this allows us to
derive 
the ground state degeneracy formula in terms of boundary
gapping conditions, which encodes more than the fusion algebra
of fractionalized quasiparticles. 
We apply our theory to Kitaev's toric code and Levin-Wen string-net models. 
We predict that the $Z_2$ toric code and $Z_2$ double-semion model (more generally, the $Z_k$ gauge theory and the $U(1)_k \times U(1)_{-k}$ non-chiral fractional quantum Hall state at even integer $k$) can be numerically and experimentally
distinguished, by measuring their boundary degeneracy on an annulus or a cylinder.  

\end{abstract}
\maketitle

\setcounter{footnote}{0} 

\noindent
\section{Introduction} 
\begin{figure}[!h]
{\includegraphics[width=.45\textwidth]{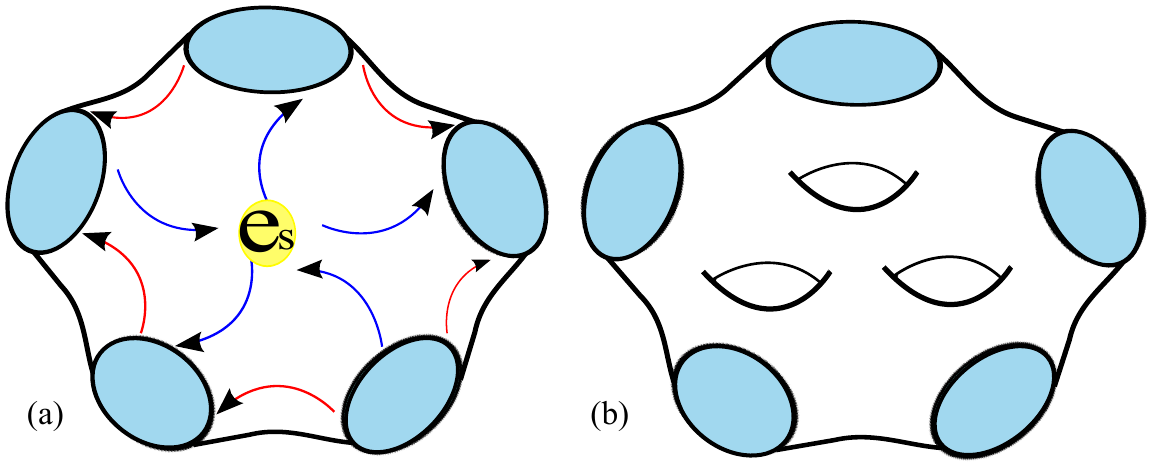}}
\caption{ Topologically ordered states on a 2D manifold with 1D boundaries: (a) 
Illustration of {\it fusion rules} and {\it total neutrality}, 
where anyons are transported from one boundary to another (red arrows), or when they fuse into physical excitations (blue arrows),
on a manifold with five boundaries.
(b) A higher genus compact surface with boundaries (thus with punctures):
a genus-3 manifold with five boundaries.}
\label{hole}
\end{figure}

Quantum many-body systems exhibit surprising new phenomena where
topological order
and the resulting fractionalization
are among the central
themes.\cite{Wen:1995qn,Wilczek:1990ik} 
Thanks to the bulk energy gap of topological order, one way to characterize topological order is through its ground state
degeneracy (GSD) on two spatial dimensional (2D) higher genus closed Riemann surface.
This GSD encodes the fusion rules of fractionalized quasiparticles and the genus
number.\cite{Wen:1989zg} However, on a 2D compact 
manifold with
boundaries (\figref{hole}), there can be gapless boundary edge modes. For
non-chiral topological orders, where the numbers of left and right moving
modes are equal, there can be interaction terms among the edge modes opening up the energy gap. 
Thus, we can ask two questions. 
First, what kinds of non-chiral topological orders provide gapped boundary edge modes, for example by introducing interaction terms?
We will show there are rules that edge modes can be fully gapped out. 
Second, when both the bulk topological order and the boundary edge modes have gapped energy spectra, we can
ask: what is the GSD of such a system?
It is the motivation of this work to understand the 
GSD for this system 
where all boundary edge excitations are gapped. 
In the following, we name this degeneracy as the \emph{boundary GSD}, to distinguish it from the \emph{bulk GSD}
of a gapped phase 
on a closed manifold without boundary.
To understand the property 
of boundary GSD 
is both of theoretical interests and of application purpose where lattice models of
topological quantum computation such as toric code\cite{Kitaev:1997wr} can be
put on space with boundaries.\cite{bk98, KitaevKong} 

In this work, we focus on topological orders in two spatial and one temporal dimensions (2+1D) without symmetry or symmetry-breaking. 
We study the topology-dependent GSD with its origin from fractionalization, not caused by symmetry-breaking.
We remark that the boundary GSD is still the GSD of the \emph{whole system including both bulk and boundaries}, 
not merely the GSD of the gapped boundary modes.
We demonstrate the boundary GSD is not simply a multiplication of
the degeneracies of all boundaries. 
In other words, the boundary GSD may not be factorizable into the degeneracy of each boundary.
We show that not only the fusion rules of fractionalized
quasiparticles (anyons) and the manifold topology, but also 
\emph{boundary gapping conditions} are the necessary data to determine the boundary GSD.
For a given bulk topological order, there are many possible types of boundary gapping conditions. 
The boundary data is not in a one-to-one correspondence or not uniquely pre-determined by the given bulk.
Specifically, the choice of boundary gapping conditions is beyond the bulk-edge correspondence.
Therefore, the boundary GSD reveals richer information than the bulk GSD.
Moreover, gluing edge modes of a compact manifold with
boundaries to form a closed manifold
, enables us to obtain the bulk
GSD from 
the boundary GSD. 

We first introduce physical concepts characterizing this boundary GSD in Sec.\ref{sec:II} and  
then rigorously derive its general formula by Chern-Simons
theory\cite{Wen:1995qn,{Witten:1988hf},{Elitzur:1989nr}}  in Sec.\ref{sec:III}.
For a concrete 
lattice realization, we implement specific
cases of our result by the $Z_2$ toric code and the string-net model in Sec.\ref{sec:IV}.\cite{Levin:2004mi}

The boundary GSD has a remarkable application to distinguish subtle differences of seemly-similar topological orders.
By measuring the boundary GSD 
on an annulus or a cylinder, 
in Sec.\ref{sec:III}, we predict the distinction between the $Z_k$ gauge theory ($Z_k$ toric code) and the $U(1)_k \times
U(1)_{-k}$ non-chiral fractional quantum Hall state at even integer $k$, despite
the two states have the same fusion algebra and the two states 
cannot be distinguished by 
the bulk GSD. For the specific $k=2$ case,
our result predicts the distinction between the $Z_2$ gauge theory ($Z_2$ toric code) and 
the twisted $Z_2$ gauge theory ($Z_2$ double-semion model) by measuring their
boundary GSD.
By using the boundary GSD as a physical observable, we can refine 
definitions of 
{\it intrinsic topological order} and {\it trivial order}, including {\it symmetry-protected topological} (SPT) {\it order},\cite{Chen:2011pg} for the case when
they have fully gapped edge modes.
Our prediction of the boundary GSD can be tested numerically by computer simulations and experimentally in the lab. 

\section{Physical Concepts} \label{sec:II}
We start by considering a topologically ordered system on a compact spatial manifold 
with boundaries, 
where each boundary have 
$N$ branches of gapless edge modes.\cite{Wen:1995qn}
Suppose the manifold has total $\eta$ boundaries, we label each boundary as $\partial_\alpha$, with $1 \leq \alpha \leq \eta$. 
Let us focus on the case that the manifold is homeomorphic to a sphere with $\eta$ punctures (\figref{hole}(a)), we will comment on cases with genus or handles (\figref{hole}(b)) later.

If 
particles condense on the boundary due to the interactions of edge modes, it can
introduce mass gap to the edge modes.  
(Note that throughout our study, we regard particles as non-fractionalized particles such as electrons,
and we regard quasiparticles as fractionalized particles such as anyons. From now on, 
we will use \emph{electron} as the synonym of \emph{particle} for the condensed matter systems.)
A set of particles can condense on the
same boundary if they do not have relative quantum fluctuation phases with each
other, thus all {\it condensed particles} are stabilized in the classical
sense.  It requires that {\it condensed particles} have relative zero braiding statistical phase 
(such as Aharonov-Bohm charge-flux braiding phase and flux-flux braiding phase), we call these 
particles with \emph{trivial braiding statistics} 
satisfying Haldane's {\it null} and {\it mutual null} conditions.\cite{h95,{Kapustin:2010hk}} 
Since electrons or particles have discrete elementary charge unit, we
label them 
as a dimension-$N$ (dim-$N$) lattice $\Gamma_e$ (here the subindex $e$ implies non-fractionalized particles such as electrons), and label 
{\it condensed
particles} as discrete lattice vectors $\ell^{\partial_\alpha}_{}$(with $\ell^{\partial_\alpha}_{} \in
\Gamma_e$) assigned to the boundary $\partial_\alpha$.  We define a {\it complete set} of
{\it condensed particles}, labeled as a lattice $\Gamma^{\partial_\alpha}_{}$, to
include all particles which have null and mutual null statistics to each other: $\Gamma^{\partial_\alpha}_{}=\{ \ell^{\partial_\alpha}_{} \}$.  

Notably 
there 
are different complete sets of condensed particles.
Assigning a {\it complete set} of {\it condensed} ({\it non-fractionalized bosonic}) 
{\it particles} to a boundary corresponds to
assigning certain type of boundary gapping conditions.  
The number of types of {\it
complete sets of condensed particles} constrains 
the number of types of \emph{boundary gapping conditions}, however, the two numbers may differ from each other (we will explore this issue in Sec.\ref{sec:bLattice_bGSD}).  

In principle, each boundary can assign its own boundary condition independently,
this assignment is not determined from the bulk information.
This is why the boundary gapping condition is \emph{beyond the bulk-edge correspondence}.
Below 
we focus on
the non-chiral orders, 
assuming all branches of edge modes can be fully gapped out. 
Later we will derive the criteria when the edge modes can be fully gapped out, at least for Abelian topological orders.

Remarkably there exists a set of {\it compatible anyons} having trivial braiding statistics
respect to the complete set of {\it condensed particles}.  
In other words, {\it compatible anyons} have mutually trivial 
braiding statistics to any elements
in the {\it complete set} of {\it condensed particles}.  For a boundary
$\partial_\alpha$, we label {\it compatible anyons} as discrete lattice vectors
$\ell^{\partial_\alpha}_{qp}$ and find all such anyons to form a {\it
complete set} labeled as $\Gamma^{\partial_\alpha}_{qp}$ with $\Gamma^{\partial_\alpha}_{qp}=\{ \ell^{\partial_\alpha}_{qp} \}$.  
Here $\Gamma^{\partial_\alpha}$ and $\Gamma^{\partial_\alpha}_{qp}$ both have the discrete Hilbert space structure as lattice.\cite{lattice-C-S}
Note that $\Gamma^{\partial_\alpha} \subseteq \Gamma^{\partial_\alpha}_{qp}$.
And $\Gamma^{\partial_\alpha}$ and $\Gamma^{\partial_\alpha}_{qp}$
have the same dimension of Hilbert space.  
If {\it compatible
anyons} can transport between different boundaries of the compact manifold, 
they must follow {\it total neutrality}: the net transport of {\it
compatible anyons} between boundaries must be balanced by the fusion 
of 
physical particles in the system  (\figref{hole}(a)), so
$\sum_\alpha \ell^{\partial_\alpha}_{qp} \in \Gamma_e$.  Transporting anyons
from boundaries to boundaries in a fractionalized manner (i.e. not in integral
electron or particle units), will result in switching the topological sectors (i.e. switching the ground states) of the
system.  Given data: $\Gamma_e, \Gamma^{\partial_\alpha},
\Gamma^{\partial_\alpha}_{qp}$, we thus derive a generic GSD formula counting the
number 
of elements in a quotient group:
\be
\GSD=\left| \frac{ \{ (\ell^{\partial_1}_{qp},\dots,\ell^{\partial_\eta}_{qp}) \mid \forall \ell^{\partial_\alpha}_{qp} \in \Gamma^{\partial_\alpha}_{qp}, \sum_\alpha \ell^{\partial_\alpha}_{qp} \in \Gamma_e  \} }
{ \{ (\ell^{\partial_1}_{},\dots,\ell^{\partial_\eta}_{}) \mid  \forall \ell^{\partial_\alpha}_{} \in  \Gamma^{\partial_\alpha}_{}  \} } \right|.  \label{gsd_L}
\ee
We derive the form of $\GSD=|L|$ with a group of discrete lattice $L$. Here $|L|$ means the number of elements in $L$, namely the order of $L$.

\section{Ground state degeneracy of Abelian topological order} \label{sec:III}
To demonstrate our above physical concepts
in a mathematically rigorous setting, let us take Abelian topological order as an example. 
It is believed that Abelian topological order can be fully classified by the $K$ matrix Abelian Chern-Simons theory.\cite{Wen:1992uk} 
For a system lives on a 2D compact manifold $\mathcal{M}$ with 1D boundaries $\partial \mathcal{M}$, edge modes of each closed boundary (homeomorphic to $S^1$) are described by 
a multiplet-chiral boson theory,\cite{Wen:1995qn} 
with the bulk action $S_{bulk}$ and the boundary action $S_{\partial}$:
\be \label{eq:Sbulk}
S_{bulk}= 
\frac{K_{IJ}}{4\pi}\int_\mathcal{M}  dt\; d^2x \; \epsilon^{\mu\nu\rho} a_{I,\mu} \partial_\nu a_{J,\rho}, 
\ee
\bea  \label{eq:Sedge}
S_{\partial}&=& \frac{1}{4\pi} \int_{\partial \mathcal{M}} dt \; dx \; K_{IJ} \partial_t \Phi_{I} \partial_x \Phi_{J} -V_{IJ}\partial_x \Phi_{I}   \partial_x \Phi_{J} \nonumber\\
&+& \int_{\partial \mathcal{M}} dt \; dx\;  \sum_{a} g_{a}  \cos(\ell_{a,I}^{} \cdot\Phi_{I}). 
\eea
Here $K_{IJ}$ and $V_{IJ}$ are symmetric integer $N \times N$ matrices, $a_{I,\mu}$ is the 1-form emergent gauge field's $I$-th component in the multiplet.
In terms of edge modes $\Phi_{I}$ with $I=1,2,\dots,N$, this means that there are $N$ branches of edge modes.
The sine-Gordon $\cos(\ell_{a,I}^{} \cdot\Phi_{I})$ is derived from a local Hermitian gapping term, 
$e^{\ti\ell_{a,I}^{} \cdot\Phi_{I}} + e^{-\ti\ell_{a,I}^{} \cdot\Phi_{I}} \propto \cos(\ell_{a,I}^{} \cdot\Phi_{I})$,
where $\ell_a^{}$ has $N$ components under index $I$ with integer coefficients.

In this work, we investigate the question how generic $g \cos(\ell_{a,I}^{} \cdot\Phi_{I})$ terms
can fully gap edge modes, by turning on large $g$ coupling interactions.
We emphasize that the perturbative relevancy/irrelevancy of $\cos(\ell_{a,I}^{} \cdot\Phi_{I})$ in the renormalization group (RG) language 
is immaterial to our large $g$ coupling limit, since there 
can have energy gap induced by non-perturbative effects at the strong interaction.
Therefore in this work we will include all possible $\ell_{a}$ terms regardless their RG relevancy.

\subsection{Canonical quantization of $K$ matrix Abelian Chern-Simons theory edge modes}
In order to understand the energy spectrum or GSD of the edge theory, we study the `quantum' theory, by 
canonical quantizing the boson field $\Phi_{I}$. 
Since $\Phi_{I}$ is the compact phase of a matter 
field, 
its bosonization has zero mode ${\phi_{0}}_{I}$ and winding momentum $P_{\phi_J}$, in addition to non-zero modes:\cite{Wen:1990se} 
\be \label{eq:mode}
\Phi_I(x) ={\phi_{0}}_{I}+K^{-1}_{IJ} P_{\phi_J} \frac{2\pi}{L}x+\ti \sum_{n\neq 0} \frac{1}{n} \alpha_{I,n} e^{-\ti n x \frac{2\pi}{L}}. 
\ee
The periodic boundary size is $L$. The conjugate momentum field of $\Phi_I(x)$ is $\Pi_{I}(x)=\frac{\delta {L}}{\delta (\partial_t \Phi_{I} )}=\frac{1}{2\pi} K_{IJ} \partial_x \Phi_{J}$. 
This yields 
the conjugation relation for zero modes:
$
[{\phi_{0}}_{I},  P_{\phi_J}]=\ti \,\delta_{IJ}
$,
and a generalized Kac-Moody algebra for non-zero modes:
$
[\alpha_{I,n} , \alpha_{J,m} ]= n K^{-1}_{IJ}\delta_{n,-m}
$.
We thus have canonical quantized fields:
$
[\Phi_I(x_1),\Pi_{J}(x_2)]= \ti \,  \delta_{IJ} \delta(x_1-x_2) 
$.\\


\subsection{Braiding Statistics and Boundary Fully Gapping Rules} \label{BFGR}
Let us 
first intuitively argue the properties of $\ell_{a}$ as {\it condensed particles} on the edge from $\cos(\ell_{a,I}^{} \cdot\Phi_{I})$ of Eq.(\ref{eq:Sedge}). 
We will leave the more rigorous justification to Sec.\ref{H_and_E}.
Let us also determine
the set of lattice spanned by the discrete integer $\ell_{a}$ vectors: $\Gamma^\partial_{}=\{\ell_{a}\}$. 
We shall name $\Gamma^\partial_{}$ as the {\it boundary gapping lattice}.
Here $a$ labels the $a$-th vector in $\Gamma^\partial$. 
From the bulk-edge correspondence, the edge condensed particles labeled by the $\ell_{a}$ vector can be mapped
to some bulk non-fractionalized particle excitations $\ell_{a}$.
It is well-known that the braiding process between two bulk excitations $\ell_{a}$ and $\ell_{b}$ of Eq.(\ref{eq:Sbulk})
causes a mutual-braiding statistical phase term to the whole wavefunction:\cite{W}
\be \label{eq:Z_ab}
 \exp[\ti \theta_{ab}]= \exp[ \ti \, 2\pi \, \ell_{a,I}^{} K^{-1}_{IJ} \ell_{b,J}^{}].
\ee
We will also denote $\ell_{a,I}^{} K^{-1}_{IJ} \ell_{b,J}^{} \equiv \ell_{a}^T K^{-1} \ell_{b}$.
On the other hand, the self-exchange process between two identical excitations $\ell_{a}$ of Eq.(\ref{eq:Sbulk})
causes a self-braiding statistical phase term to the whole wavefunction:\cite{W}
\be \label{eq:Z_aa}
 \exp[\ti \theta_{aa}/2]= \exp[ \ti \, \pi \, \ell_{a,I}^{} K^{-1}_{IJ} \ell_{a,J}^{}].
\ee
Without any global symmetry constraint, then any gapping term is allowed. 
Below we argue what are the list of properties that the gapping term satisfies to fully gap the edge modes: 

\noindent
(i) Bosonic self-statistics: 
$\forall \ell_a \in \Gamma^\partial_{}$, $\ell_{a,I}^{} K^{-1}_{IJ} \ell_{a,J}^{} \in 2 \mathbb{Z}$ even integers. This means that the self-statistics of $\ell_{a}$ is bosonic, with a multiple $2\pi$ phase. \\
(ii) Local:  
$\forall \ell_a^{}$, $\ell_b^{} \in \Gamma^\partial_{},\; \ell_{a,I}^{} K^{-1}_{IJ} \ell_{b,J}^{} \in \mathbb{Z}$ integers. 
Winding one $\ell_{a}$ around another $\ell_{b}$ yields a bosonic phase, namely a multiple $2\pi$ statistical phase. 
The bosonic statistics can be viewed as the \emph{local} condition.\\
(iii) Localizing condensate at the classical value without being eliminated by self or mutual quantum fluctuation:
${\forall  \ell_a^{}, \ell_b^{} \in \Gamma^\partial_{},\; \ell_{a,I}^{} K^{-1}_{IJ} \ell_{b,J}^{} =0} $, 
so that $\mathbf{Z}_{\text{statistics}} \sim  \exp[\ti \theta_{ab}] =1$,
the condensation is stabilized and survives in the classical sense.\\
(iv) For the $\cos(\ell_{a,I}^{} \cdot\Phi_{I})$ term, ${\ell}_a^{}$ must be excitations of 
non-fractionalized particle degrees of freedom, since it lives on the `physical' boundary, so ${\ell}_a^{} \in \Gamma_e$ lattice, where 
\be
\Gamma_e=\{ \sum_J c_J K_{IJ} \mid  c_J \in \mathbb{Z} \}.
\ee
This rule imposes an integer charge $ q_I K^{-1}_{IJ} {\ell}^{}_{a,J}$ in the bulk, 
and an integer charge $Q_{I}=\int^L_0 \frac{1}{2\pi} \partial_x \Phi_I dx= K^{-1}_{IJ} P_{\phi_{J}} = K^{-1}_{IJ} {\ell}^{}_{a,J}$ for each branch of edge mode $I$ on the boundary.  Here $q_I$ is the charge vector coupling to an external field $A_\mu$ of gauge or global symmetry, by adding $A_\mu q_I J^\mu_I$ to the ${S}_{bulk}$, 
which corresponds to $q_I A^\mu \partial_\mu \Phi_I$ in the ${S}_{\partial}$.\\
(v) Completeness: we define $\Gamma^\partial_{}$ is a complete set, by including
every possible term $\ell_c$ that has the self null braiding statistics and has the mutually null braiding statistics respect to all the elements $\ell_a \in \Gamma^\partial_{}$.
Namely, mathematically we have $\forall \ell_c^{} \in \Gamma_e$, if $\ell_c^{T} K^{-1} \ell_c^{}=0$ and $\ell_c^{T} K^{-1} \ell_a^{}=0$ for $\forall \ell_a^{} \in \Gamma^\partial_{}$, then $\ell_c^{} \in \Gamma^\partial_{}$ must be true. Otherwise $\Gamma^\partial_{}$ is not complete. \\
(vi) The system is non-chiral. We require the same number of left moving modes and right moving modes to fully gap out the edge modes.

In Sec.\ref{H_and_E} we will use the bulk braiding statistics property of $\ell_{a}$ to determine the gapped edge stability caused by $\cos(\ell_{a,I}^{} \cdot\Phi_{I})$ of Eq.(\ref{eq:Sedge}).
We leave a derivation that these properties above are \emph{sufficient conditions} in Sec.\ref{H_and_E}. 

Indeed the above rules (i)(ii)(iii)(iv)(v)(vi) can be simplified to a set of rules which we call {Boundary Fully Gapping Rules}. 

\subsubsection{Boundary Fully Gapping Rules} \label{sec:BFGR}

For an Abelian topological order described by a bulk Chern-Simons theory of Eq.(\ref{eq:Sbulk}) and a boundary theory of Eq.(\ref{eq:Sedge}), we can
add a set of proper interaction terms $\cos(\ell_{a,I}^{} \cdot\Phi_{I})$ on the boundary to gap out the edge modes. 
We will term that the Boundary Fully Gapping Rules, which summarize all the above rules (i)(ii)(iii)(iv)(v)(vi) to determine the gapping term $\ell_a \in \Gamma^\partial_{}$. 
Here $\ell_a$ is some integer vector, namely for every component $\ell_{a,I} \in \mathbb{Z}$. 
The $\Gamma^\partial_{}$ satisfies:\\
 
\noindent
({\bf1}) Null and mutual null conditions:\cite{h95} ${\forall  \ell_a^{}, \ell_b^{} \in \Gamma^\partial_{}}$, ${\ell_{a,I}^{} K^{-1}_{IJ} \ell_{b,J}^{} =0} $. 
This implies self statistics and mutual statistics are bosonic, and the excitation is local. Localized fields are 
not eliminated by self or mutual quantum fluctuations,
so the condensation survives in the classical sense.\\

\noindent
({\bf2}) The dimensions of the lattice $\Gamma^\partial_{}$ is $N/2$, where $N$ must be an even integer. 
Namely, the Chern-Simons lattice $\Gamma^\partial_{}$ assigned to a boundary $\partial$ is spanned by $N/2$ linear independent vectors $\ell_a$.
Mathematically, we write $\Gamma^{\partial}_{}=\{ \underset{a=1, 2, \dots, {N/2}}{\sum}   I^{}_{a} {\ell^{}_{a,I}} \mid\;I^{}_{a} \in \mathbb{Z}\}$. 
\\



\noindent
({\bf 3}) The system is non-chiral.
The signature of $K$ matrix (defined as the number of positive eigenvalues $-$ the number of negative eigenvalues, as $n_L-n_R$) must be zero. 
The non-chiral edge modes implies a measurable observable, the thermal Hall conductance,\cite{Kane_Fisher} to be zero $\kappa_{xy}= (n_L-n_R)\frac{\pi^2 k_B^2}{3 h}T=0$.
Again, $N=n_L+n_R$ is even.\\

There is an extra rule, which will be important later when we try to reproduce the bulk GSD from the boundary GSD:\\

\noindent
({\bf4}) `Physical' excitation: ${\ell}_a^{} \in \Gamma_e=\{\sum_J c_J K_{IJ} \mid  c_J \in \mathbb{Z} \}$. 
Namely, ${\ell}_a^{}$ is an excitation of non-fractionalized particle 
degree of freedom, 
since it lives on the `physical' boundary.\\

Our justification of {Boundary Fully Gapping Rules} as the \emph{sufficient conditions} to gap the edge is left to Sec.\ref{H_and_E}.

\subsubsection{Comments}

Here are some comments for the above rules.
Since any linear combinations of $\ell_a^{} \in \Gamma_e$ still satisfy ({\bf1})({\bf2})({\bf3}), 
we can regard 
$ \Gamma^\partial_{}$ as an {\it infinite discrete lattice group} generated by some basis vectors $\ell_a^{}$.

%
Physically, the rule ({\bf 3}) excludes some violating examples such as odd rank (denoted as rk) $K$ matrix with 
the chiral central charge $c_-=c_L-{c}_R\neq 0$ or the thermal Hall conductance $\kappa_{xy} \neq 0$, which universally has 
gapless chiral edge modes.
For instance, the dim-$1$ boundary gapping lattice: $\{ n(A,B,C) \mid n \in \mathbb{Z} \}$ of $K_{3\times 3}=\diag(1,1,-1)$, with 
$A^2+B^2-C^2=0$, satisfies the rules ({\bf 1})({\bf 2}), but cannot fully gap out chiral edge modes.

Moreover, from the above rules
we find $\sqrt{|\det K|}$ belongs to a positive integer, namely 
\be \label{eq:sq}
\sqrt{|\det K|} \in \mathbb{N}^+.
\ee
We will show an explitict calculation for the $K_{2\times 2}$-matrix Chern-Simons theory in the Appendix \ref{SecK22}. 
One can generlize our result to a higher rank $K$-matrix Chern-Simons theory.\\


\subsection{Hamiltonian and Energy Gap} \label{H_and_E}

Here we will justify the Boundary Fully Gapping Rules in Sec.\ref{BFGR} is \emph{sufficient} to fully gap the edge modes. Our approach is to
explicitly calculate the mass gap for the zero energy mode and its higher excitations. We will show that 
\emph{if} the Boundary Fully Gapping Rules hold, there are \emph{stable mass gaps} for all edge modes. 

We consider the even-rank symmetric $K$ matrix, satisfying the rule ({\bf 3}), so the non-chiral system with even number of edge modes can potentially be gappable.

To determine the mass gap of the boundary modes, and to examine the gap in the large system size limit $L\to \infty$, we will take the large $g$ coupling limit of the Hamiltonian: 
$
-g_{a} \int_0^{L} \; dx\; \cos(\ell_{a,I}^{} \cdot\Phi_{I}) \to \frac{1}{2}g_{a}(\ell_{a,I}^{} \cdot\Phi_{I})^2 L
$.
By exactly diagonalizing the quadratic Hamiltonian,
\be
H \simeq (\int^L_0 dx\; V_{IJ}\partial_x \Phi_{I}   \partial_x \Phi_{J} )+\frac{1}{2} \sum_a g_a (\ell_{a,I}^{} \cdot\Phi_{I})^2 L +\dots, 
\ee
with a $\Phi$ mode expansion Eq.(\ref{eq:mode}), we obtain the energy spectra from its eigenvalues.
We realize that:\\
{$\bullet$ \bf Remark 1}: \emph{If} we include all the interaction terms allowed by {\bf Boundary Full Gapping Rules}, we can turn on the energy gap of zero modes ($n=0$) 
as well as the Fourier modes (non-zero modes $n\neq 0$). 
The energy spectrum is in the form of
\be \label{eq:ap-mass-gap-st}
E_n= \big( \sqrt{ \Delta^2 + \# (\frac{2\pi n}{L})^2 } + \dots \big),
\ee
where $\Delta$ is the mass gap. Here $\#$ means some numerical factor.
We emphasize the energy of Fourier modes ($n\neq 0$) behaves towards zero modes 
at long wave-length low energy limit ($L \to \infty$). Such spectra become continuous at $L \to \infty$ limit, which is the expected energy behavior.

\noindent
{$\bullet$ \bf Remark 2}: \emph{If} we include the \emph{incompatible} interaction term, 
e.g. $\ell_a$ and $\ell'$ where $\ell_a^{T} K^{-1} \ell' \neq 0$, 
while the interaction terms contain $\sum_a g_a \cos(\ell_{a} \cdot \Phi) +g' \cos(\ell_{}' \cdot \Phi)$, we obtain the \emph{unstable} energy spectrum:
\be \label{eq:Enunstable}
E_n= \big( \sqrt{ \Delta_m^2 + \# (\frac{2\pi n}{L})^2+ \sum_{a} \# g_{a}  \,g' (\frac{L}{n})^2 +\dots } + \dots \big).
\ee
The energy spectra exhibits an \emph{instability} of the system, because at low energy limit ($L \to \infty$), the spectra become discontinuous (from $n=0$ to $n \neq 0$) and jump to infinity as long as 
there are incompatible cosine terms (i.e. $g_a \cdot g' \neq 0$). 
The dangerous behavior of $(L/n)^2$ implies the quadratic expansion analysis may not describe the full physics.
In that case, the dangerous behavior invalidates localizing of $\Phi$ field at a minimum. This 
invalidates the energy gap, and the \emph{unstable} system potentially seeks to become \emph{gapless phases}.

\noindent
{$\bullet$ \bf Remark 3}: 
%
We provide an alternative way to study the energy gap stability.
We include 
the full cosine interaction term for the lowest energy states, namely the zero and winding modes:
\be \label{eq:cos}
\cos(\ell_{a,I}^{} \cdot\Phi_{I}) \to  
\cos(\ell_{a,I}^{} \cdot ({\phi_{0}}_{I}+K^{-1}_{IJ} P_{\phi_J} \frac{2\pi}{L}x) ).
\ee
The stability of the energy gap can be understood from \emph{under what criteria} we can safely expand the cosine term to extract the leading quadratic terms 
by only keeping the zero modes, namely $\cos(\ell_{a,I}^{} \cdot\Phi_{I}) \simeq 1 - \frac{1}{2}(\ell_{a,I}^{} \cdot\phi_{0I})^2 +\dots$.
The naive reason is the following: if one does not decouple the winding mode $P_{\phi_J}$ term, there is a complicated $x$ dependence in $P_{\phi_J} \frac{2\pi}{L}x$ along the $x$ integration.
The \emph{non-commuting} algebra $[{\phi_{0}}_{I},  P_{\phi_J}]=\ti \delta_{IJ}$ results in the challenge for this cosine expansion.
This challenge can be resolved by requiring $\ell_{a,I}^{}  {\phi_{0}}_{I}$ and $\ell_{a,I'}^{}  K^{-1}_{I'J} P_{\phi_J}$ \emph{commute} 
in Eq.(\ref{eq:cos}),
\bea
[\ell_{a,I}^{}  {\phi_{0}}_{I}, \;\ell_{a,I'}^{}  K^{-1}_{I'J} P_{\phi_J}] &=&\ell_{a,I}^{} K^{-1}_{I'J} \ell_{a,I'}^{}  \; (\ti\delta_{IJ}) \nonumber\\
&=&(\ell_{a,J}^{} K^{-1}_{I'J} \ell_{a,I'}^{} )(\ti)=0.\;\;\;\;\;\;
\eea  
In fact this is the {Boundary Full Gapping Rule ({\bf 1})} for the self null statistics --- the trivial 
self statistics rule among the interaction gapping terms. 
We can interpret that there is \emph{no quantum fluctuation} destabilize the semi-classical particle condensation.
With this \emph{commuting criterion}, 
we can safely expand Eq.(\ref{eq:cos}) by the trigonometric identity as
\bea
&&\cos(\ell_{a,I}^{} {\phi_{0}}_{I}) \cos(\ell_{a,I}^{} K^{-1}_{IJ} P_{\phi_J} \frac{2\pi}{L}x) \nonumber \\
&&-\sin(\ell_{a,I}^{} {\phi_{0}}_{I}) \sin(\ell_{a,I}^{} K^{-1}_{IJ} P_{\phi_J} \frac{2\pi}{L}x). 
\eea
Then we integrate over the circumference $L$. 
First, we notice that $\ell_{a,I}^{} K^{-1}_{IJ} P_{\phi_J}$ takes integer values due to $\ell_{a,I}^{} \in \Gamma_e$ and $P_{\phi_J} \in \mathbb{Z}$.
Further we notice that 
due to the periodicity of both $\cos(\dots x)$ and $\sin(\dots x)$ in the region $[0,L)$, so both $x$-integrations over $[0,L)$ vanish. However, 
the exception is $\ell_{a,I}^{} \cdot K^{-1}_{IJ} P_{\phi_J} =0$, then $\cos(\ell_{a,I}^{} K^{-1}_{IJ} P_{\phi_J} \frac{2\pi}{L}x)=1$.
We derive:
\be \label{eq:int_cos}
 g_{a} \int_0^{L}dx\; \text{Eq}.(\ref{eq:cos})=g_{a} L \; \cos(\ell_{a,I}^{} \cdot {\phi_{0}}_{I}) \delta_{(\ell_{a,I}^{} \cdot K^{-1}_{IJ} P_{\phi_J} ,0)}.
\ee
The Kronecker-delta function $\delta_{(\ell_{a,I}^{} \cdot K^{-1}_{IJ} P_{\phi_J} ,0)}=1$ 
indicates that there is a nonzero contribution if and only if $\ell_{a,I}^{} \cdot K^{-1}_{IJ} P_{\phi_J} =0$. 

So far we have shown 
that when the self-null braiding statistics $\ell^T K^{-1} \ell=0$ is true, we have the desired cosine potential expansion via the zero mode quadratic expansion at the large $g_a$ coupling,
$  g_{a} \int_0^{L}dx \cos(\ell_{a,I}^{} \cdot\Phi_{I}) \simeq  - g_{a} L \frac{1}{2}(\ell_{a,I}^{} \cdot\phi_{0I})^2 +\dots$.
If we include not enough gapping terms (less than $N/2$ terms), we cannot fully gap all edge modes. 
On the other hand, if we include more than the {Boundary Full Gapping Rules} (more than $N/2$ terms with incompatible terms), there is a disastrous behavior in the spectrum (see {\bf Remark 2}).
We need to include the mutual-null braiding statistics $\ell^T_a K^{-1} \ell_b=0$ so that the energy gap is stable.

The quadratic Hamiltonian includes both the kinetic and the leading-order of the potential terms:
\be \label{eq:H}
\frac{(2\pi)^2}{4\pi L}  V_{IJ} K^{-1}_{I l_1} K^{-1}_{J l_2} P_{\phi_{l_1}} P_{\phi_{l_2}} +\sum_a g_{a} L  \frac{1}{2}(\ell_{a,I}^{} \cdot\phi_{0I})^2
\ee
By solving the quadratic simple harmonic oscillators, we can show the nonzero energy gaps of zero modes.
The mass matrix can be properly diagonalized, since there are only conjugate variables $\phi_{0I},P_{\phi,J}$ in the quadratic order.
The energy gap is of the order one finite gap, independent of the system size $L$,
\be
\Delta_{} \simeq O(\sqrt{2\pi\, g_a \ell_{a,l_1} \ell_{a,l_2} V_{IJ} K^{-1}_{I l_1} K^{-1}_{J l_2} }). 
\ee
In the diagonalized basis of the Hamiltonian Eq.(\ref{eq:H}), the energy gap $\Delta_{I}$ has the component $I$-dependence. 

More precisely, we find the dimension of independent gapping terms $\Gamma^\partial_{}=\{\ell_a\}$ must be $N/2$, namely satisfying {Boundary Full Gapping Rules ({\bf 2})}.
The number of left and right moving modes must be the same, namely satisfying the non-chiral criterion in {Boundary Full Gapping Rules ({\bf 3})}.
To summarize, by calculating the stability of energy gap, we have thus 
demonstrated that the {Boundary Full Gapping Rules ({\bf 1})({\bf 2})({\bf 3})} are \emph{sufficient} to ensure that the energy gap is stable at large $g$ coupling.

Due to the periodicity of ${\phi_{0}}_{}$, its conjugate variable $P_{\phi}$ forms a discrete quantized lattice. %
This is consistent with the discrete Hilbert space of the ground states, forming the \emph{Chern-Simons quantized lattice} detailed in Sec.\ref{sec:Hilbert}.
We will apply this idea to count the ground state degeneracy of the Chern-Simons theory on a closed manifold or a compact manifold with gapped boundaries in the next Sec.\ref{sec:bLattice_bGSD}.
The {Boundary Full Gapping Rules ({\bf 4}) will be required for the boundary GSD and the bulk GSD in Sec.\ref{sec:bLattice_bGSD}.


\subsection{Hilbert Space} \label{sec:Hilbert}

Since ${\phi_{0}}_{}$ is periodic, so 
$P_{\phi}$ forms a discrete lattice.
We now impose 
the rule ({\bf 4}), 
so $\cos(\ell_{a,I}^{} \cdot {\phi_{0}}_{I})$ 
are hopping terms along {\it condensed particle} vector $\ell_{a,I}^{}$ in sublattice of $\Gamma^e$ in the $P_{\phi}$ lattice.    
We will show that rule ({\bf 4}) is essential to derive the bulk GSD by computing the boundary GSD under gluing the boundaries in Sec.\ref{sec:bLattice_bGSD}.

Let $P^{qp}_{\phi}$ represents some {\it compatible anyon} $\ell_{qp}$ which is mutual null to {\it condensed particles} $\ell$ by $\ell_{}^{T} K^{-1}_{} P^{qp}_{\phi}=\ell_{}^{T} K^{-1} \ell_{qp} =0$.
By 
the rule ({\bf1}), thus it means that the compatible anyon $\ell_{qp}$ parallels along 
some $\ell_{}^{}$ vector. However, $\ell_{qp}$  lives on the quasiparticle lattice, i.e. the unit integer lattice of the $P_{\phi}$ lattice. So $\ell_{qp} $ is parametrized by $\frac{1}{|\gcd(\ell_a^{})|} \ell_{a,J}^{}$, with the greatest common divisor defined as $|\gcd(\ell_a)| \equiv  \gcd(|\ell_{a,1}|,|\ell_{a,2}|,\dots,|\ell_{a,N}|)$. 

Now let us consider the Hilbert space of ground states in terms of $P_{\phi}$ lattice.
For the Hilbert space of ground states, we will neglect the kinetic term $H_{kin}=\frac{(2\pi)^2}{4\pi L}  V_{IJ} K^{-1}_{I l1} K^{-1}_{J l2} P_{\phi_{l1}} P_{\phi_{l2}}$ of the order $O(1/L)$ as $L \to \infty$. 
Recall we label the $\alpha$-th boundary of a compact spatial manifold with $\eta$ punctures as $\partial_\alpha$, where $\alpha =1, \dots, \eta$.
Note that $a$ is the index for $a$-th $\ell$ vector: $\ell^{\partial_\alpha}_{a} \in \Gamma^{\partial_\alpha}_{}$. 
If we choose the proper basis $\ell$ vector, based on the rule ({\bf 2}), we have $a=1,\dots,N/2$. 
For the $\alpha$-th boundary $\partial_\alpha$,
a {\it complete set} of {\it condensed particles} forms the {\it boundary gapping lattice}:
\be
\Gamma^{\partial_\alpha}_{}=\{ \sum_{a=1,\dots, N/2}  I^{\partial_\alpha}_{a} {\ell^{\partial_\alpha}_{a,I}} \mid\;I^{\partial_\alpha}_{a} \in \mathbb{Z}\}. 
\ee
Recall $I$ is the $I$-th branch of $K_{N \times N}$ matrix, $I=1,\dots, N$. 

A {\it complete set} of {\it compatible anyon} vectors $\ell_{qp}$ forms the Hilbert space of the winding mode $P_{\phi}$ lattice: 
\bea
\Gamma^{\partial_\alpha}_{qp}=\{\ell^{\partial_\alpha}_{qp,I}  \} =\{ \sum_{a=1,\dots, N/2}   j^{\partial_\alpha}_{a} \frac{ \ell^{\partial_\alpha}_{a,I} }{ |\gcd(\ell^{\partial_\alpha}_{a})|  } \mid \;  j^{\partial_\alpha}_{a} \in \mathbb{Z} \}, \;\;\;\;\;\;\;
\eea
or simply the {\it anyon hopping lattice}. 
 Note $\Gamma^{\partial_\alpha}_{}$, $\Gamma^{\partial_\alpha}_{qp}$ are {\it infinite Abelian discrete lattice group}.
Anyon fusion rules and the {\it total neutrality} condition essentially means the bulk physical charge excitation can {\it fuse} from or {\it split} to multiple anyon charges. 
The rules constrain the set of $j^{\partial_\alpha}_{a}$ values to be limited on the $\Gamma_e$ lattice.

To be more precise mathematically, the anyon fusion rules and the total neutrality condition 
constrain the direct sum\cite{direct_sum} of the {\it anyon hopping lattice} $\Gamma^{\partial_\alpha}_{qp}$, with $\alpha=1,\dots, \eta$ over all $\eta$ boundaries, must be on the $\Gamma_e$ lattice.
We define such a constrained anyon hopping lattice as $L_{qp \bigcap e}$:
%
\bea
L_{qp \bigcap e}\equiv &\{&\bigoplus_{\alpha=1}^\eta \sum_{a=1}^{N/2}  j^{\partial_\alpha}_{a} \frac{\ell^{\partial_\alpha}_{a,I} }{ |\gcd(\ell^{\partial_\alpha}_{a})|} \; \mid \;  \forall j^{\partial_\alpha}_{a}\in\mathbb{Z},\;   \exists \;c_J\in\mathbb{Z},\nonumber \\
 && \sum_{\alpha=1}^\eta  \sum_{a=1}^{N/2} j^{\partial_\alpha}_{a} \frac{ \ell^{\partial_\alpha}_{a,I}}{ |\gcd(\ell^{\partial_\alpha}_{a})|}=\sum_{J=1}^N c_J K_{IJ}\}.
 \eea  

\subsubsection{Hilbert Space of Ground States}

Now we focus on further understanding the ground state eigenvectors and their Hilbert space.
At large $g$ coupling, we can view the interaction term $g_{a}  \cos(\ell_{a,I}^{} \cdot\Phi_{I})$ as 
a potential term pinning down the $\Phi_I$ field at the minimum of the potential energy.

The periodicity of ${\phi_{0}} \sim {\phi_{0}} + 2 \pi $ gives the quantization of its conjugate variable $P_{\phi} \in \mathbb{Z}$.
In terms of operator forms, by the commutation relation $[\hat{\phi_{0}}, \hat{P_{\phi}}]=i$, we find\\
\bea
&&e^{-i n \hat{{\phi_{0}}}} \hat{{P_{\phi}}} e^{i n \hat{\phi}} = \hat{{P_{\phi}}} +n,\\
&&e ^{i \hat{{P_{\phi}}} s} | {\phi_{0}} \rangle = \; | {\phi_{0}} - s \rangle, \\
&&e ^{i n \hat{{\phi_{0}}}} | {P_{\phi}} \rangle = | {P_{\phi}}+n\rangle, \label{eq:P_hopping}
\eea
up to some scaling factors. For the ground state concerning the zero modes and winding modes, we can express 
 its lowest energy Hamiltonian at the large $g$ limit containing Eq.(\ref{eq:int_cos}) in terms of the well-defined operators $e^{i \hat{\phi_{0}}}$ and $\hat{{P_{\phi}}}$:
 \bea
&& H_0=-g_{a} L \; \cos(\ell_{a,I}^{} \cdot {\hat{\phi_{0}}}_{I}) \delta_{(\ell_{a,I}^{} \cdot K^{-1}_{IJ} \hat{P}_{\phi_J} ,0)}\\
&& = -\frac{g_{a}}{2} L \; (e^{i\ell_{a,I}^{} \cdot {\hat{\phi_{0}}}_{I}}+e^{-i\ell_{a,I}^{} \cdot {\hat{\phi_{0}}}_{I}}) \delta_{(\ell_{a,I}^{} \cdot K^{-1}_{IJ} \hat{P}_{\phi_J} ,0)}.
 \eea
 
There are two ways to think about the ground states. The first way is that viewing the ground state from the Hilbert space of all possible zero modes 
${\phi_{0}}_I$: $\mathcal{H}=\{ | {\phi_{0}}_I \rangle \}$. 
In this way, a typical ground state is \emph{pinned down} at a minimum of the cosine potential:
\be \label{eq:phi_gs}
| {\phi_{0}}_{I} \rangle.
\ee
The second way to think about the ground state is that viewing it from the Hilbert space of winding modes ${P}_{\phi_J}$ only.
The full Hilbert space is 
\be
\mathcal{H}=\{ | {P_{\phi_J}} \rangle \}, \text{ where } {P_{\phi_J}}\in \mathbb{Z},
\ee
up to some extra constraints due to the cosine potential (hopping terms), such as the delta function constraint in Eq.(\ref{eq:int_cos}).
In this dual description, the ground state will be hopping around on the ${P}_{\phi_J}$ lattice. 
From Eq.(\ref{eq:P_hopping}), we learn that $e^{i\ell_{a,I}^{} \cdot {\hat{\phi_{0}}}_{I}}$ will forward hop $|{P}_{\phi_I} \rangle$ along the $\ell_{a,I}^{}$ vector with a distance $|\ell_{a,I}^{}|$.
Similarly, $e^{-i\ell_{a,I}^{} \cdot {\hat{\phi_{0}}}_{I}}$ will backward hop $|{P}_{\phi_I} \rangle$ along the $-\ell_{a,I}^{}$ vector  with a distance $|\ell_{a,I}^{}|$.
So this ground state $| {\phi_{0}}_{I} \rangle$ can be also expressed in $| {P_{\phi_J}} \rangle$ basis:
\be \label{eq:P_phi_gs}
\sum_{ \underset{n_a\in \mathbb{Z}, \; \forall a}{{P_{\phi_J}} =n_a \ell_{a,J},}} | {P_{\phi_J}} \rangle  \cdot \langle{P_{\phi_J}} | {\phi_{0}}_{I} \rangle,
\ee
which is the Fourier transformation of Eq.(\ref{eq:phi_gs}).
We find that using the ${P}_{\phi_J}$ lattice Hilbert space has its convenience, better than
the ${\phi_{0}}_{I}$-Hilbert space,
 when there are multiple boundaries assigned with multiple gapped boundary conditions.
 In Sec.\ref{sec:transport}, we describe a physical way to switch topological sectors, thus switch ground states,
by transporting anyons. 
In Sec.\ref{sec:bLattice_bGSD}, we will derive the GSD formula in the ${P}_{\phi_J}$ lattice.

\subsection{Transport between Ground State Sectors: Flux Insertion Argument and Experimental Test on Boundary Types} \label{sec:transport}

Let us consider the anyon transport in the simplest topology --- an annulus or a cylinder.
Consider an artificial-designed gauge field or an external gauge field (such as electromagnetic field) $A$ coupled to topologically ordered states by a charge vector $q_I$.
An adiabatic flux insertion $\Delta \Phi_B$ through the cylinder induces an electric field ${E}_x$ through the Faraday effect. 
The electric field ${E}_x$ causes a perpendicular current $J_y$ flows to the boundary through the Hall effect. 
We can precisely calculate the induced current $J$ from 
the bulk term $J_J^\mu=-q_I \frac{e}{2 \pi} K^{-1}_{IJ} \frac{c}{\hbar}\epsilon^{\mu \nu \rho} \partial_\nu A_\rho$, so
\bea
&&q_I \Delta \Phi_B=-q_I \int dt \int\vec{E} \cdot d\vec{l} \nonumber \\
&&=-\frac{2 \pi }{e} K_{IJ}\hbar\int J_{y,J} dt dx =-\frac{2 \pi }{e} K_{IJ}\frac{\hbar}{e} Q_J. \nonumber 
\eea
Here $Q_J$ is the charge condensed on the edge of the cylinder.
On the other hand, the edge dynamics affects winding modes by 
\be
Q_I=\int J^0_{\partial,I} dx=-\int \frac{e}{2\pi}  \partial_x \Phi_I dx =-e K^{-1}_{IJ} P_{\phi,J}.
\ee 
Combine the above two effects, we obtain: 
\be \label{eq:anyon transport}
q_I \Delta \Phi_B/(\frac{h}{e})=\Delta P_{\phi,I}.
\ee
An adiabatic flux change $\Delta \Phi_B$ induces the anyon transport from one boundary to another,
 and switches the winding mode by $\Delta P_\phi$.
Apply Eq.(\ref{eq:anyon transport}) to Eq.(\ref{eq:P_phi_gs}), we learn that, as long as $|\Delta P_{\phi,I}|$ is smaller than the hopping amplitude $|\ell_{a,I}^{}|$ of
Eq.(\ref{eq:P_phi_gs}), we will shift the ground state to another sector.
More explicitly, we will shift a ground state from:
$\sum_{ \underset{n_a\in \mathbb{Z}, \; \forall a}{{P_{\phi_J}} =n_a \ell_{a,J},}} | {P_{\phi_J}} \rangle  \cdot \langle{P_{\phi_J}} | {\phi_{0}}_{I} \rangle$
to another ground state
$\sum_{ \underset{n_a\in \mathbb{Z}, \; \forall a}{{P_{\phi_J}} =n_a \ell_{a,J}+\Delta P_{\phi,J},}} | {P_{\phi_J}} \rangle  \cdot \langle{P_{\phi_J}} | {\phi_{0}}_{I} \rangle$. 

By counting the number of all distinct ground states (here within this ${P_{\phi_J}}$-hopping lattice), we can determine the GSD.

\subsection{Boundary Gapping Lattice, Boundary Gapping Condition, and Ground State Degeneracy} \label{sec:bLattice_bGSD}
 
\subsubsection{Ground State Degeneracy} 
 
The GSD counts the number of topological sectors distinguished by the fractionalized anyons transport between boundaries. (See the way of transport in Sec.\ref{sec:transport}.)  
The direct sum of condensed particle lattice $\bigoplus_{\alpha=1}^\eta \Gamma^{\partial_\alpha}$ obviously satisfies 
the anyon fusion rules and the {total neutrality} condition, 
therefore the lattice $L_{qp \bigcap e}$ contains the lattice $\bigoplus_{\alpha=1}^\eta \Gamma^{\partial_\alpha}$. 
More precisely, we know that $\bigoplus_{\alpha=1}^\eta \Gamma^{\partial_\alpha}$ is a normal subgroup of $ L_{qp \bigcap e}$. 
Therefore, given the input data $K$ and $\Gamma^{\partial_\alpha}$ (which are sufficient to determine $\Gamma^{\partial_\alpha}_{qp}$),
we derive %
the GSD is the number of elements in a {\it quotient finite Abelian group}:
\bea
\GSD =\left|  {\frac{  L_{qp \bigcap e}  }{  \bigoplus_{\alpha=1}^{\eta} \Gamma^{\partial_\alpha}_{}   } } \right|,    \label{gsd}
\eea
analogous to Eq.\;(\ref{gsd_L}).
Interestingly the GSD formula Eq.\;(\ref{gsd}) works for both closed manifolds or compact manifolds with boundaries. 
By gluing the boundaries of a compact manifold, we can enlarge the original $K_{N\times N}$ matrix to a $K_{2N\times 2N}$ matrix of glued edge modes and 
create $N$ scattering channels to fully gap out all edge modes. 
For a genus $g$ Riemann surface with $\eta'$ punctures (\figref{hole}(b)), we start with a number of $g$ cylinders drilled with extra punctures,\cite{append1} use Eq.\;(\ref{gsd}) to account for glued boundaries  which contributes at most a $|\det K|^g$ factor, and redefine particle hopping lattices $L_{qp \bigcap e}$ and $\bigoplus_{\alpha'} \Gamma^{\partial_{\alpha'}}_{} $ only for unglued boundaries ($1\leq \alpha' \leq \eta'$), we obtain
\be
\GSD \leq |\det K|^g \cdot  \left|  \frac{ L_{qp \bigcap e}}{ \bigoplus^{\eta'}_{\alpha'=1} \Gamma^{\partial_{\alpha'}}_{}  } \right|, \label{gsdG}
\ee
if the system has no symmetry-breaking.
For a genus $g$ Riemann surface ($\eta'=0$), 
Eq.\;(\ref{gsdG}) becomes 
$
\GSD \leq |\det K|^g \nonumber
$.
The inequalities are due to 
different choices of gapping conditions for 
glued boundaries.\cite{WW} 
Further details can be found in Appendix \ref{sec:genus_g_R}.

Below we apply our algorithm to 
the generic rank-2 $K_{2\times 2}$ matrix case. (The explicit calculation is saved to Appendix \ref{SecK22}.) 
From Eq.(\ref{eq:sq}), to fully gap out the edge modes of $K_{2\times 2}$-Chern-Simons theory requires $\det K=-k^2$ with an integer $k$. 
Take a cylinder with two gapped boundaries ${\partial_1}$ and ${\partial_2}$ as an example (equivalently a sphere with two punctures), 
Eq.\;(\ref{gsd}) shows $\GSD=\sqrt{|\det K|}=k$ when boundary gapping conditions on two edges are the same;  
namely, we find that $\GSD=\sqrt{|\det K|}=k$ when the two boundary gapping lattices satisfy $\Gamma^{\partial_1}_{}=\Gamma^{\partial_2}_{}$.
However, the GSD on a cylinder yields $\GSD\leq \sqrt{|\det K|}= k$ 
when boundary gapping lattices on two edges are different: $\Gamma^{\partial_1}_{}\neq \Gamma^{\partial_2}_{}$.  


For specific examples, we take the $Z_k$ gauge theory ($Z_k$ toric code) formulated by a 
$K_{Z_k}=\bigl( {\begin{smallmatrix} 
0 &k \\
k & 0 
\end{smallmatrix}} \bigl)$ Chern-Simons theory and take the 
$U(1)_k \times U(1)_{-k}$ non-chiral fractional quantum Hall state   
formulated by a 
 $K_{\diag,k}=\bigl( {\begin{smallmatrix}
k &0 \\
0 & -k
\end{smallmatrix}} \bigl)$ Chern-Simons theory. 
By computing the GSD on a cylinder with different boundary gaping lattices (i.e. $\Gamma^{\partial_1}_{}\neq \Gamma^{\partial_2}_{}$), 
we find $K_{Z_k}$ has $\GSD=1$, while $K_{\diag,k}$ has $\GSD=1$ for odd $k$ but $\GSD=2$ for even $k$. See Table \ref{table1}.
\begin{table}[h]
\begin{tabular}{c|C||C|C }
      \multicolumn{2}{c||}{GSD} & K_{Z_k} & K_{\diag,k} \\ \hline\hline
      Boundary & \Gamma^{\partial_1}_{} \neq \Gamma^{\partial_2}_{} & 1& \text{1 ($k\in \odd$) or 2 ($k\in \even$)}     \\ \cline{2-4}
      \;GSD & \Gamma^{\partial_1}_{} = \Gamma^{\partial_2}_{} & k & k\\ \hline
      \multicolumn{2}{c||}{Bulk GSD}& k^2 &  k^2\\
    \end{tabular}
\caption{Boundary GSD on a cylinder with two gapped edges and bulk GSD on a 2-torus for 
the $Z_k$ gauge theory (with $K_{Z_k}$) 
and the $U(1)_k \times U(1)_{-k}$ non-chiral fractional quantum hall state (with $K_{\diag,k}$).
}
\label{table1}
\end{table}

\begin{table}[h]
    \begin{tabular}{c|C||C|C }
      \multicolumn{2}{c||}{ GSD} & K_{Z_2}: \text{toric code} & K_{\diag,2}: \text{double-semion} \\ \hline\hline
      Boundary & \Gamma^{\partial_1}_{} \neq \Gamma^{\partial_2}_{} & 1& 2    \\ \cline{2-4}
      \;GSD & \Gamma^{\partial_1}_{} = \Gamma^{\partial_2}_{} & 2 & 2\\ \hline
      \multicolumn{2}{c||}{Bulk GSD}& 2^2 &  2^2\\
    \end{tabular}
\caption{Boundary GSD on a cylinder with two gapped edges and bulk GSD on a 2-torus 
for the $Z_2$ toric code ($Z_2$ gauge theory with $K_{Z_2}$) and the $Z_2$ double-semion model (twisted $Z_2$ gauge theory with $K_{\diag,2}$). }
\label{table1-Z2}
\end{table}

Table \ref{table1} shows a  
new surprise. We predict a distinction between two classes of topological orders: $Z_k$ gauge theory (with $K_{Z_k}$) 
and $U(1)_k \times U(1)_{-k}$ non-chiral fractional quantum hall state (with $K_{\diag,k}$) at even integer $k$ by simply measuring their boundary GSD on a cylinder.
We can take $k=2$ case in Table \ref{table1} for the more familiar lattice model examples: the $Z_2$ toric code ($Z_2$ gauge theory with $K_{Z_2}$)
and the $Z_2$ double-semion model (twisted $Z_2$ gauge theory with $K_{\diag,2}$), shown in Table \ref{table1-Z2}.
By computing the GSD on a cylinder with different gapped boundaries (i.e. $\Gamma^{\partial_1}_{}\neq \Gamma^{\partial_2}_{}$), 
we find the $Z_2$ toric code has $\GSD=1$, while the $Z_2$ double-semion model has $\GSD=2$.

 \subsubsection{Boundary Gapping Lattice v.s. Boundary Gapping Condition} \label{sec:bGL_bGC}

In Sec.\ref{sec:II}, we mention that the \emph{boundary gapping lattice} $\Gamma^{\partial}$ derived from the Boundary Fully Gapping Rules,
is associated to certain \emph{boundary gapping condition}. However, their relation is not in a one-to-one correspondence.
In this subsection, we will address the precise relation between the boundary gapping lattice $\Gamma^{\partial}$ and the boundary gapping condition.
See Table \ref{table:bGSD_TO} for our explicit computations of the boundary gapping conditions and the number of types of boundary gapping conditions, $\mathcal{N}^\partial_g$.

On one hand, the boundary gapping lattice may \emph{over-count} the number of boundary gapping conditions.
For example, for the $Z_2$ {double-semion} model described by 
$K_{\diag,2}=\bigl( {\begin{smallmatrix}
2 &0 \\
0 & -2
\end{smallmatrix}} \bigl)$, we find
two boundary gapping lattices $\Gamma^\partial$ and $\Gamma^{\partial'}$: \\
\noindent
(1) $\Gamma^\partial=\{ n \, \ell^\partial = n(2,2) \mid n \in \mathbb{Z} \}$ with compatible anyons 
$\Gamma^\partial_{qp}=\{n \, \ell^\partial_{qp} =  n(1,1) \mid n \in \mathbb{Z} \}$.\\  
(2) $\Gamma^{\partial'}=\{ n \, \ell^{\partial'} = n \, \ell^{\partial'}=n(2,-2) \mid n \in \mathbb{Z} \}$  with compatible anyons 
$\Gamma^{\partial'}_{qp}=\{ n \, \ell^{\partial'}_{qp} = n(1,-1) \mid n \in \mathbb{Z} \}$.
Even though the boundary gapping lattices  of $\Gamma^\partial_{qp}$ and $\Gamma^{\partial'}_{qp}$ look different, 
but their lattice structures are transformable to each other via identifying the bulk non-fractionalized particles. 
Namely, the lattice structure of both $\Gamma^\partial_{qp}$ and $\Gamma^{\partial'}_{qp}$ are transformable to each other via the particle lattice vectors $\Gamma_e$:
\be \label{eq:identify gapping lattices}
\ell^\partial_{qp,I} =\ell^{\partial'}_{qp,I} + \sum_J c_J K_{IJ}
\ee
Since we have $\ell^\partial_{qp,I}=n(1,1)=n(1,-1)-n(0,-2) =\ell^{\partial'}_{qp,I} + \sum_J c_J K_{IJ}$.
Thus, for $Z_2$ double-semion, there is only one boundary gapping condition: $\mathcal{N}^\partial_g=1$, 
see Table \ref{table:bGSD_TO}.
More generally,  for two sets of boundary gapping lattices $\Gamma^\partial$ and $\Gamma^{\partial '}$ with corresponding anyon hopping lattice $\Gamma^{\partial}_{qp}$ and $\Gamma^{\partial'}_{qp}$,
we know the two sets give rise to the identical boundary gapping condition if we can identify them via Eq.(\ref{eq:identify gapping lattices}).
We can label each boundary gapping condition by the distinct set of compatible anyons identified via Eq.(\ref{eq:identify gapping lattices}).

On the other hand, the boundary gapping lattice may \emph{undercount} the number of boundary gapping conditions.
In Sec.\ref{BFGR}, we use the \emph{null condition}:\cite{h95} the braiding statistical phase to be zero, in order to demonstrate the gapped edge and the estimated mass gap in Sec.\ref{H_and_E}.
However, the \emph{null condition} may be too strong: Boundary Fully Gapping Rules are proven to be sufficient but may not be necessary.
Indeed, if we loosen the mutual-braiding statistics to:
\be \label{eq:braiding_1}
\ell_{a,I}^{} K^{-1}_{IJ} \ell_{b,J}^{} \in \mathbb{Z},
\ee
and loosen the self-braiding statistics to:
\be \label{eq:braiding_2}
\ell_{a,I}^{} K^{-1}_{IJ} \ell_{a,J}^{} \in \left\{ \begin{array}{ll} 2\mathbb{Z}, \text{ for bosonic systems.}\\ \mathbb{Z}, \text{ for fermionic systems.} \end{array}\right.,
\ee
we can still define the statical phases of Eq.(\ref{eq:Z_ab}) and Eq.(\ref{eq:Z_aa}) to be trivial: a bosonic system obtaining a $+1$ phase,
and a fermionic system obtaining a $\pm 1$ phase. 

An example that the boundary gapping lattice \emph{under-counts} the number of boundary gapping conditions is 
the $Z_4$ gauge theory described by $K_{Z_4}=\bigl( {\begin{smallmatrix}
0 & 4 \\
4 & 0
\end{smallmatrix}} \bigl)$.
We find
two boundary gapping lattices $\Gamma^\partial$ and $\Gamma^{\partial'}$: \\
\noindent
(1) $\Gamma^\partial=\{ n \, \ell^\partial = n(4,0) \mid n \in \mathbb{Z} \}$ with compatible anyons 
$\Gamma^\partial_{qp}=\{n \, \ell^\partial_{qp} =  n(1,0) \mid n \in \mathbb{Z} \}$.\\  
(2) $\Gamma^{\partial'}=\{ n \, \ell^{\partial'} = n \, \ell^{\partial'}=n(0,4) \mid n \in \mathbb{Z} \}$  with compatible anyons 
$\Gamma^{\partial'}_{qp}=\{ n \, \ell^{\partial'}_{qp} = n(0,1) \mid n \in \mathbb{Z} \}$.\\
However, there is another set of compatible anyons which satisfies the trivial statistical rules Eq.(\ref{eq:braiding_1}) and Eq.(\ref{eq:braiding_2}): $\{ (2,0),(0,2),(2,2),\dots \}$.\\
In this case, we may include an extra boundary gapping lattice outside of Boundary Fully Gapping Rules:\\
(3) 
$\Gamma^{\partial''}=\{ n(4,0) +m(0,4)\mid n, m \in \mathbb{Z} \}$ and 
$\Gamma^{\partial''}_{qp}=\{ n(2,0) +m(0,2)\mid n, m \in \mathbb{Z} \}$.\\
Thus, for $Z_4$ gauge theory, there are three boundary gapping condition: $\mathcal{N}^\partial_g=3$.
We can still use the formula Eq.(\ref{gsd}) to calculate
 the boundary GSD on the cylinder with two edges assigned different boundary gapping conditions, the boundary GSD can be $1,2,4$, see Table \ref{table:bGSD_TO}.
Here the boundary GSD as 1 and 4 are already captured by Table \ref{table1}.
The $\GSD=4$ is due to the same boundary types on two sides of a cylinder.
The $\GSD=1$ is due to the different boundary types $\Gamma^{\partial}$ and $\Gamma^{\partial'}$ on two sides of a cylinder.
The $\GSD=2$ occurs when the different boundary types contain $\Gamma^{\partial''}$ on one side, and contain $\Gamma^{\partial}$ or $\Gamma^{\partial'}$ on the other side of a cylinder.

More generally, the notion of the compatible anyons with trivial braiding statistics of Eq.(\ref{eq:braiding_1}) and Eq.(\ref{eq:braiding_2}) 
is termed \emph{Lagrangian subgroup}, studied independently by Ref.\onlinecite{Kapustin:2010hk}, \onlinecite{Levin:2013gaa,{Barkeshli:2013yta}}.

\begin{widetext}
\begin{center}
\begin{table}[h]
\begin{tabular}{|c||c| c| c| c|}
\hline   
Bosonic Topological Orders & $\mathcal{N}^\partial_g$ & Boundary Gapping Conditions & GSD on a torus $=|\det K|$ &  GSD on an annulus \\  
\hline
\hline 
{
$\begin{array}{ll} K^{}_{Z_2}= 
{\begin{pmatrix} 
0 & 2 \\
2 & 0 
\end{pmatrix}}\\
Z_2 \text{ toric code}
\end{array}
  $}& 2 & $ \left. 
 \begin{array}{ll} 
\{ (1,0),(2,0),\dots \},\\
\{(0,1),(0,2),\dots \}
  \end{array}  \right.$ & 4 & 1, 2 \\ 
\hline
{
$\begin{array}{ll} K^{}_{\diag,2}= 
{\begin{pmatrix} 
2 & 0 \\
0 & -2 
\end{pmatrix}}\\
Z_2 \text{ double-semion}
\end{array}
  $}
& 1 & $ 
\{ (1,1),(2,2),\dots \}
$ & 4 & 2 \\ 
\hline
{
$\begin{array}{ll} K^{}_{Z_3}= 
{\begin{pmatrix} 
0 & 3 \\
3 & 0 
\end{pmatrix}}\\
\text{} Z_3 \text{ gauge theory}
\end{array}
  $}
& 2 & $ \left. 
 \begin{array}{ll} 
\{ (1,0), (2,0), (3,0),\dots \},\\
\{ (0,1),(0,2),(0,3),\dots \}
  \end{array}  \right.$ & 9 & 1, 3  \\ 
  \hline
{
$\begin{array}{ll} K^{}_{Z_4}= 
{\begin{pmatrix} 
0 & 4 \\
4 & 0 
\end{pmatrix}}\\
Z_4 \text{ gauge theory}
\end{array}
  $}& 3 & $ \left. 
 \begin{array}{lll} 
\{ (1,0),(2,0),(3,0),\dots \},\\
\{(0,1),(0,2),(0,3),\dots \},\\
\{ (2,0),(0,2),(2,2),\dots \}
  \end{array}  \right.$ & 16 & 1, 2, 4 \\ 
\hline
{
$\begin{array}{ll} K^{}_{\diag,4}= 
{\begin{pmatrix} 
4 & 0 \\
0 & -4 
\end{pmatrix}}\\
\text{} U(1)_4 \times U(1)_{-4}  \text{ FQH}
\end{array}
  $}& 2 & $ \left. 
 \begin{array}{ll} 
\{ (1,1),(2,2),(3,3),\dots \},\\
\{ (1,3),(2,2),(3,1),\dots \}
  \end{array}  \right.$ & 16 &  2, 4 \\ 
\hline

\hline
Fermionic Topological Orders & $\mathcal{N}^\partial_g$ & Boundary Gapping Conditions &  GSD on a torus $=|\det K|$ & GSD on an annulus\\ 
\hline
\hline
{
$\begin{array}{ll} K^{}_{\diag,3}= 
{\begin{pmatrix} 
3 & 0 \\
0 & -3 
\end{pmatrix}}\\
\text{} U(1)_3 \times U(1)_{-3}  \text{ FQH}
\end{array}
  $}
& 2 & 
$ \left. 
 \begin{array}{ll} 
\{ (1,1),(2,2),(3,3),\dots \},\\
\{ (1,2),(2,1),(3,3),\dots \}
  \end{array}  \right.$
  & 9 & 1, 3 \\ 
\hline
\end{tabular}
\caption{In the first column, we list down some bosonic and fermionic topological orders and their $K$-matrices in Chern-Simons theory. 
Non-fractionalized particles of bosonic topological orders can have only bosonic statistics, but
non-fractionalized particles of fermionic topological orders can have fermionic statistics.
In the second column, we list down their number of types of  boundary gapping conditions $\mathcal{N}^\partial_g$.
In the third column, we list down their boundary gapping conditions in terms of a set of compatible and condensable anyons with trivial braiding statistics.
In the fourth column, we list down their bulk GSD$=|\det K|$ on a closed manifold 2-torus.
In the fifth column, we list down their boundary GSD on an annulus (or a cylinder) with all various types of boundary gapping conditions on two edges.  
The $U(1)_k \times U(1)_{-k}$ FQH means the doubled layer chiral and anti-chiral fractional quantum hall (FQH) states combine to be a non-chiral topological order.}
\label{table:bGSD_TO}
\end{table}
\end{center}
\end{widetext}

In summary, as we exactly solve the number of types of boundary gapping lattices, 
we find that for $\rank(K)=2$, we obtain two boundary gapping lattices.
However, when we consider boundary gapping conditions, we
apply the identification Eq.(\ref{eq:identify gapping lattices})
and the trivial statistical rules Eq.(\ref{eq:braiding_1}) and Eq.(\ref{eq:braiding_2}),
we obtain a list of number of types of boundary gapping conditions $\mathcal{N}^\partial_g$
in Table \ref{table:bGSD_TO}, where $\mathcal{N}^\partial_g \neq 2$ in general.

For a $K$-matrix Chern-Simons theory with $\rank(K) \geq 4$, 
we find there can be infinite number of sets of \emph{boundary gapping lattices}.
For example, as $K_{4\times 4}=\diag(1,1,-1,-1)$, 
one can find a dim-2 boundary gapping lattice, $\{ n(A,B,C,0),m(0,C,B,A) \mid n,m \in \mathbb{Z}, A^2+B^2-C^2=0 \}$. 
Different sets of $A,B,C$ give different lattices.
However, when we consider \emph{boundary gapping conditions}, we need to
apply the identification Eq.(\ref{eq:identify gapping lattices})
and the trivial statistical rules Eq.(\ref{eq:braiding_1}) and Eq.(\ref{eq:braiding_2}).
We find that there are only two representative sets, labeled by:\\
(1) $\{ n (1,0,1,0) + m (0,1,0,1) \mid n, m \in \mathbb{Z} \}$,\\
(2) $\{ n (1,0,0,1) + m (0,1,1,0) \mid n, m \in \mathbb{Z} \}$.\\
The boundary gapping lattices $\{ n(A,B,C,0),m(0,C,B,A) \mid n,m \in \mathbb{Z}, A^2+B^2-C^2=0 \}$ 
can be always reduced to these two sets.
However, this two sets can be identified via Eq.(\ref{eq:identify gapping lattices}), since $K$ matrix has unit integers in each column.
So there is \emph{only one} boundary gapping condition $\mathcal{N}^\partial_g=1$.
See Appendix \ref{sec:Ng} for a discussion on the bosonic and fermionic trivial orders with $|\det K|=1$ have \emph{only one} boundary gapping condition, $\mathcal{N}^\partial_g=1$.

For other Abelian topological orders described by $K$-matrix Chern-Simons theories, 
there can be finite numbers of boundary gapping conditions. 
The most important message for the types of boundary gapping conditions is that:
We should view \emph{the set of compatible anyons} as \emph{the condensation of particles or anyons} with trivial braiding statistics of Eq.(\ref{eq:braiding_1}) and Eq.(\ref{eq:braiding_2}), which 
defines the \emph{boundary gapping conditions} (the third column in Table \ref{table:bGSD_TO}).
    
\noindent
\section{Examples of boundary GSD: Mutual Chern-Simons theory, ${Z}_k$ topological order, toric code and string-net model} \label{sec:IV}
We now take the ${Z_k}$ gauge theory example with a $K_{Z_k}$-matrix Chern-Simons theory to
demonstrate our understanding of two types of GSD on a cylinder with gapped boundaries in physical pictures. 
By checking all the fusion and braiding properties of quasiparticle excitations, we know that the ${Z_k}$ gauge theory and the $K_{Z_k}=\bigl( {\begin{smallmatrix} 
0 &k \\
k & 0 
\end{smallmatrix}} \bigl)$ Chern-Simons theory are indeed equivalent to the mutual Chern-Simons theory:
$\frac{k}{2\pi}\int  dt\; d^2x \; \epsilon^{\mu\nu\rho} a_{1,\mu} \partial_\nu a_{2,\rho}$.  
All these describe the so-called ${Z}_k$ topological order.
%
\begin{figure}[!h]
{\includegraphics[width=.45\textwidth]{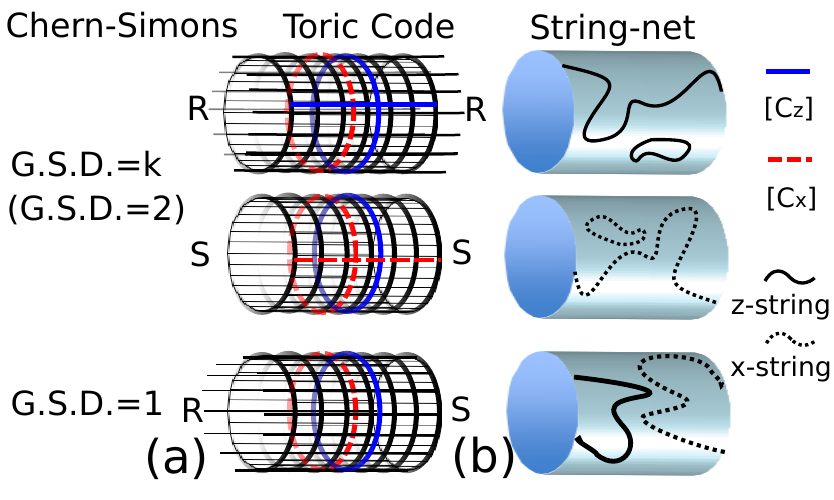}}
\caption{(a) The same boundary conditions on two ends of a cylinder allow a pair of 
cycles $[c_{x}],[c_{z}]$ of a qubit, 
thus $\GSD=2$. Different boundary conditions do not, thus $\GSD=1$.
(b) The same boundary conditions allow z- or x-strings connect two boundaries. Different boundary conditions do not.}
\label{toric_stringnet}
\end{figure}

When $k=2$, it realizes $Z_2$ toric code 
with a Hamiltonian $H_0=-\sum_v A_v - \sum_p B_p$ on a square lattice.\cite{{Kitaev:1997wr},Kou:2008rn} 
Here the convention is that the vertex operator $A_v=\prod \sigma^x$ goes around four neighbor links of a vertex 
and the plaquette operator $B_p=\prod \sigma^z$ goes around four neighbor links of a plaquette,
with Pauli matrices $\sigma^x$ and $\sigma^z$.
Since the Kitaev's toric code is well-known, the reader can consult other details defined in Ref.\onlinecite{Kitaev:1997wr}.
There are two types of gapped boundaries\cite{bk98} on a cylinder (\figref{toric_stringnet}(a)): First, the $x$ boundary 
(or the rough boundary, denoted as R in FIG.\ref{toric_stringnet}) where $z$-string charge $e$-charge condenses. Second, 
the $z$ boundary (or the smooth boundary, denoted as S in FIG.\ref{toric_stringnet}) where $x$-string ``charge'' $m$-flux condenses.\cite{Kitaev:1997wr} 
We can determine 
the GSD by counting the degree of freedom of the code subspace: 
the number of the qubits --- the number of the independent stabilizers.
For $\Gamma^{\partial_1}_{} = \Gamma^{\partial_2}_{}$, we have the same number of qubits and stabilizers, with one extra constraint $\prod_{\text{all sites}} B_p=1$ for two $x$-boundaries 
(similarly, $\prod_{\text{all sites}} A_v=1$ for two $z$-boundaries). 
This leaves $1$ free qubit, thus 
$\GSD=2^1=2$.
 For $\Gamma^{\partial_1}_{} \neq \Gamma^{\partial_2}_{}$, still the same number of qubits and stabilizers, but has no extra constraint. This leaves no free qubits, thus
 $\GSD=2^0=1$.

We can also count the number of independent logical operators (\figref{toric_stringnet}(a)) in the homology class, with the string-net picture (\figref{toric_stringnet}(b))  in mind. There are two cycles $[c_{x_1}],[c_{z_1}]$ winding around the compact direction of a cylinder. 
If both gapped boundaries of a cylinder are $x$-boundaries, we only have $z$-string connecting two edges: the cycle $[c_{z_2}]$. 
If both gapped boundaries of a cylinder are $z$-boundaries, we only have $x$-string (dual string) connecting two edges: the cycle $[c_{x_2}]$. 
We can define the qubit algebra 
by using the generators of $[c_{x_1}],[c_{z_2}]$ in the first case and 
by using the generators of $[c_{x_2}],[c_{z_1}]$ in the second case.
Cycles of either case can define the algebra $\sigma^x,\sigma^y,\sigma^z$ of a qubit, 
so $\GSD=2$. 
If gapped boundaries of a cylinder are different (one is $x$-boundary, the other is $z$-boundary),
we have no string connecting two edges: there is no nontrivial cycle, which yields no nontrivial Lie algebra, and $\GSD=1$.

Let us use the string-net picture to view the ground state sectors and the GSD.
For both $x$-boundaries ($z$-boundaries), one ground state has even number of strings (dual strings), the other ground state has odd number of strings (dual strings), connecting two edges;
so again we obtain $\GSD=2$.
On the other hand, if the boundaries are different on two sides of the cylinder, no cycle is allowed in the non-compact direction, no string and no dual string can connect two edges, so $\GSD=1$. 

Generally, for a $Z_k$ gauge theory (as a level $k$ doubled model) on the compact orientable spatial manifold $\mathcal{M}$ without boundaries or with gapped boundaries, 
without symmetry and without symmetry-breaking,
we obtain its GSD is bounded by the order of the first homology group $H_1(\mathcal{M},Z_k)$ of $\mathcal{M}$ with $Z_k$ coefficient,\cite{fm98}
or equivalently the $k$ to the power of 
the $1$st Betti number ${b_1(\mathcal{M})}$, namely\cite{WW}  
\be
\GSD \leq |H_1(\mathcal{M},Z_k)| =|\bigoplus_{k=1}^{b_1(\mathcal{M})} Z_k| = k^{b_1(\mathcal{M})}
\ee


\section{Topological Order and Trivial Order }
Now let us ask a fundamental 
question: what is topological order? 
For a 2D bulk state with non-chiral fully-gapped boundary modes, 
we realize the original definition of degenerated ground states on a higher genus Riemann surface\cite{{Wen:1995qn},{Wen:1989zg}} can be transplanted to 
degenerated ground states on an annulus with two boundaries. 
We define: 
\emph{For a non-chiral fully-gapped system on the boundary and in the bulk, without symmetry and without symmetry-breaking,
the state is an intrinsic topological order if 
it has degenerated ground states (
at least for certain boundary gapping conditions) on an annulus}. 
For Abelian topological orders described by Abelian Chern-Simons theory, the ``\emph{at least}'' statement 
is due to $\GSD\leq \sqrt{|\det K|}$, 
only the same boundary condition on two sides of an annulus gives $\GSD=\sqrt{|\det K|}$. 

Similarly, without symmetry and without symmetry-breaking,
\emph{the state is trivial order without topological order, if it has a unique ground state on an annulus with two gapped boundaries, for any boundary gapping condition}. 

On the other hand, with symmetry, for 2D SPT order,  the edge modes cannot be gapped without breaking the symmetry. If the edge modes of SPT order are fully gapped by breaking the symmetry,
 then there can be $\GSD \geq 1$ on an annulus due to the symmetry-breaking degenerate ground states.

\section{Conclusion}
We introduce the new concept of the ground state degeneracy of topological order on a manifold with a gapped bulk and gapped boundaries for the energy spectrum ---
finite number of ground states are isolated from the energetic excitations.
This concept is termed as the boundary degeneracy. 
We compute the boundary GSD formula for 2D Abelian topologically ordered states by using a $K_{N\times N}$ matrix Abelian Chern-Simon theory. 
We show that Boundary Fully Gapping Rules are sufficient to fully gap the edge modes, and we examine
the low energy Hamiltonian and Hilbert space, boundary gapping lattices, and the number of types of boundary gapping conditions $\mathcal{N}^\partial_g$. 

To have fully gapped boundaries for the Chern-Simons theory requires that the rank $N$ is even and non-chiral. 
This reflects the gapped boundary properties of string-net models or \emph{Quantum Doubled} models including toric codes.\cite{KitaevKong,BSW2011}
We compute the number of boundary gapping conditions $\mathcal{N}^\partial_g$ for several topological orders in Table \ref{table:bGSD_TO}.
We confirm the two boundary gapping conditions of toric code.\cite{bk98,fm98,BSW2011} 

We show that counting the boundary GSD can 
reproduce the bulk GSD by gluing the boundaries of a manifold and gapping edge modes on both boundaries.
However, we find there are more types of the boundary GSD instead of a unique bulk GSD, 
depending on the types of boundary gapping conditions. 
A remarkable example is the $Z_2$ toric code and $Z_2$ double-semion model 
(more generally, the $Z_k$ gauge theory and the $U(1)_k \times U(1)_{-k}$ non-chiral fractional quantum Hall state, 
described by $K_{Z_k}$ and $K_{\diag,k}$ Chern-Simons theory at even integer $k$): 
Though the GSD of both states on a closed genus $g$ surface are indistinguishable $(\GSD=k^{2g})$,
their boundary GSD on a cylinder are different.
The example is especially surprising because both states have the same $(\mathbb{Z}_k)^2$ fusion algebra. 
This means the fusion algebra alone cannot determine boundary GSD. 
In the 
category theory language,\cite{KitaevKong} the model of the unitary fusion category $\mathcal{C}$ shows that (in Table \ref{table2})
there can be many different $\mathcal{C},\mathcal{D}$ types realizing the same 
 monoidal center $Z(\mathcal{C})=Z(\mathcal{D})$.
In other words, there are many gapped boundaries correspond to the same gapped bulk.

\begin{table}[h]
\begin{tabular}{c||c}
Physics & Category \\ \hline
Bulk excitation &  objects in unitary modular category   \\
(anyons) & $Z(\mathcal{C})$ (monoidal center of $\mathcal{C}$)\\ \hline
Boundary type & the set of equivalent classes $\{\mathcal{C},\mathcal{D},\dots\}$\\   
&of unitary fusion category
 \end{tabular}
\caption{Dictionary between physics and category}
\label{table2}
\end{table}

\noindent 
%
Finally, our definitions of {\it topological orders} not only deepen understanding of topological GSD, 
but also ease the experimental platform with only an annulus topology instead of higher genus surfaces.
For future research directions,
it will be interesting to 
realize the boundary GSD and the number of boundary gapping conditions $\mathcal{N}^\partial_g$ without using the $K$-matrix Chern-Simons theory, 
which is restricted only to Abelian topological order.
Physical concepts which we introduced in Sec.\ref{sec:II} should still hold universally. 
Other than the {\it fusion rules} and the {\it total neutrality} condition, whether {\it braiding rule} can explicitly enter into the GSD formula?\cite{braiding}
%
%
These shall inspire generalizing Eq.\;(\ref{gsd}) to the boundary GSD of non-Abelian topological orders.


It will be illuminating to have more predictions based on our theory, 
as well as experimental realizations of boundary types. 
One approach 
is described in Sec.\ref{sec:transport}: the flux insertion
through an annulus or a cylinder.  An adiabatic flux change $\Delta \Phi_B$ induces the anyon transport from one boundary to another by 
$\Delta \Phi_B/({h}/{e})=\Delta P_\phi$.  
The change of winding mode  
$\Delta P_\phi$ can switch the ground state sector. 
It will be interesting to see how the same type of boundary gapping conditions allows this effect (a unit flux insertion shifts the ground state to another topological sector, with the total number of sectors as 
$\GSD=\sqrt{|\det K|}$), while different types of boundary gapping conditions restrain this effect \emph{dynamically} ($\GSD<\sqrt{|\det K|}$). 
For the same type of gapped boundaries, there are $\sqrt{|\det K|}$ sectors labeled by $\Delta P_\phi({\bmod} \;\sqrt{|\det K|})$, 
the $\sqrt{|\det K|}$ units of flux bring the state back to the original sector. 
For different types of gapped boundaries,  
we had shown $\GSD<\sqrt{|\det K|}$ (such as $\GSD=1$). 
This motivates an interesting question if one inserts flux into the cylinder, what dynamical effect, which repulses anyons transporting from one gapped edge to the other, will be detected. 
The detection of this dynamical effect can guide experiments to distinguish boundary types, namely the boundary gapping conditions.

\begin{acknowledgments}
JW acknowledges Maissam Barkeshli,  Liang Kong, Tian Lan, John McGreevy,
William Witczak-Krempa and Lucy Zhang for comments. 
This work is supported by NSF Grant No. DMR-1005541, NSFC 11074140, and
NSFC 11274192.  
It is also supported by the BMO
Financial Group and the John Templeton Foundation.
Research at Perimeter Institute is supported by the
Government of Canada through Industry Canada and by the Province of Ontario
through the Ministry of Research. 
\end{acknowledgments}


Note add:
JW thanks Tian Lan for collaborating a related work Ref.\onlinecite{Lan:2014uaa} and a discussion on Table \ref{table:bGSD_TO}.
After the completion of this work (arXiv:1212.4863), 
we become aware that Ref.\onlinecite{{Levin:2013gaa},{Barkeshli:2013yta}} later have
independently studied a similar criteria of gapped boundaries for Abelian topological orders. 
Anton Kapustin has independently derived a related result of the boundary GSD.\cite{Kapustin:2013nva}

\begin{widetext}
\end{widetext}

\appendix

{\bf Appendix}

In Appendix, we 
demonstrate our algorithm and GSD formula  Eq.\;(\ref{gsd}) for a generic rank-2 $K$ matrix in Sec.\ref{SecK22}.
We also give an example why fusion algebra alone does not provide enough information to determine the boundary GSD from the bulk-edge correspondence viewpoint. 
In Sec.\ref{sec:genus_g_R}, we outline the gluing technique to derive the GSD formula Eq.\;(\ref{gsdG}) of a compact manifold with genus.
In Sec.\ref{sec:Ng}, we comment more about the number of boundary types, $\mathcal{N}^\partial_g$. 

\section{Analysis on a $K_{2 \times 2}$ Chern-Simons Theory} \label{SecK22}

Here we work through a rank-2 $K$ matrix Chern-Simons theory example, to demonstrate our generic algorithm in the main text. We will derive
its low energy Hamiltonian, Hilbert space, boundary GSD formula, and the number of types of boundary gapping lattices.
Generally we write a rank-2 $K$ matrix as $K_{2 \times 2}=\bigl( {\begin{smallmatrix} 
k_1 &k_3 \\
k_3 & k_2 
 \end{smallmatrix} } \bigl)  \equiv\bigl( {\begin{smallmatrix} 
k_1 &k_3 \\
k_3 & (k_3^2-p^2)/k_1 \
 \end{smallmatrix} } \bigl)$. In order to fully gap out edge modes, we find that: 
First, the edge modes need to be non-chiral ($K_{2 \times 2}$ with the equal number of positive and negative eigenvalues), so $\det K<0$. 
Second, the $|\det K|$ needs to be an integer $p$ square, $\det K=-p^2$.

We find two independent sets of the allowed gapping lattices 
$\Gamma^\partial_{}=\{ n {\ell_{a,I}^{}} | n \in \mathbb{Z} \}$ and 
$\Gamma^{\partial'}_{}=\{ n' {\ell_{a,I}^{}}' | n' \in \mathbb{Z} \}$ 
satisfying gapping rules ({\bf1})({\bf2})({\bf3})({\bf4})
at $\rank(K)=2$,
with
\bea
&&n\ell_{a,I}^{}=n(\ell_{a,1}^{},\ell_{a,2}^{})= \frac{n\; p}{|\gcd(k_1,k_3+p)|} (k_1,k_3+p), \;\;\;\;\;\;\;\;\label{eq:l}  \\
&&n' {\ell'_{a,I}}=n'({\ell'_{a,1}},{\ell'_{a,2}})= \frac{n'\;p}{|\gcd(k_3+p,k_2)|} (k_3+p,k_2). \;\;\;\;\;\;\;\;\label{eq:l'}
\eea
Here $|\gcd(k,l)|$ stands for finding the greatest common divisor in $|k|,|l|$ and taking its absolute value. 
If $k$ (or $l$) is zero, we define $|\gcd(k,l)|$ is the other value $|l|$ (or $|k|$). 
Here $n,n' \in \mathbb{Z}$ are allowed if no other symmetry constrains its values.

Now we will take two specific topology, 
a disk (a sphere with 1 puncture) and a cylinder (a sphere with 2 punctures), as examples of manifolds with boundaries. 
For $K_{2 \times 2}$, the Hilbert space of edge modes on the disk is:\cite{Wen:1990se}
\be
\mathcal{H}_{disk}= \mathcal{H}^{1,2}_{KM}  \otimes \mathcal{H}_{P_{\phi_1}} \otimes \mathcal{H}_{P_{\phi_2}}.
\ee
(If $K_{2 \times 2}$ is diagonal, then $\mathcal{H}^{1,2}_{KM} =\mathcal{H}^1_{KM} \otimes \mathcal{H}^2_{KM}$.) 
The Hilbert spaces of edge modes on the cylinder is
\bea
&&\mathcal{H}_{cylinder}= \bigoplus_{j_A} \bigoplus_{j_B} (\mathcal{H}^{top}_{disk} \otimes \mathcal{H}^{bottom}_{disk} )^{(j_A,j_B)}\\
&=&\mathcal{H}^{top}_{disk} \otimes \mathcal{H}^{bottom}_{disk} \otimes  \mathcal{H}_{gl}  \\
&=&\mathcal{H}^{top,1,2}_{KM}  \otimes \mathcal{H}^{bottom,3,4}_{KM}  \otimes \mathcal{H}_{P^{top}_{\phi_1},P^{bottom}_{\phi_3}}  \nonumber
\otimes \mathcal{H}_{P^{top}_{\phi_2},P^{bottom}_{\phi_4}}
\eea
$\mathcal{H}_{KM}$ stands for the Hilbert space of nonzero Fourier mode part with Kac-Moody algebra.
We label the low energy Hilbert space by winding mode ${P_{\phi}}$, 
which can be regarded as a discrete lattice because of $\Phi_I(x)$ periodicity.
The $1,2,3,4$ indices stand for the component (branch index) of $\Phi_I(x)$.
Because the bulk cylinder provides channels 
connecting edge modes of two boundaries, so fractionalized quasiparticles (here Abelian anyons) can be transported from one edge to the other. 
$\mathcal{H}_{gl}$ contains fractional sectors $| j_A,j_B \rangle$, the 1st branch $j_A$ runs between the top ($P^{top}_{\phi_1}$) and the bottom ($P^{bottom}_{\phi_3}$),
the 2nd branch $j_B$ runs between the top ($P^{top}_{\phi_2}$) and the bottom ($P^{bottom}_{\phi_4}$).

Let us explicitly show that edge modes with 
these gapping terms Eq.(\ref{eq:l}) and (\ref{eq:l'}) 
have a finite energy gap above the ground states at a large system size $L$ and a large coupling $g$. 
Without losing generality, take $V=\bigl( {\begin{smallmatrix} v_1 &v_2 \\
  v_2 & v_1 \\
\end{smallmatrix}} \bigl)
$ and a gapping term $(\ell_{a,1}^{},\ell_{a,2}^{})=\frac{p}{|\gcd(k_1,k_3+p)|} (k_1,k_3+p) \in \Gamma^\partial_{}$, and diagonalize the Hamiltonian, 
\be
H \simeq (\int^L_0 dx\; V_{IJ}\partial_x \Phi_{I}   \partial_x \Phi_{J} )+\frac{1}{2}g (\ell_{a,I}^{} \cdot\Phi_{I})^2 L 
\ee
we find energy eigenvalues:
\be
E_{1,2}(n)=\sqrt{ {\Delta^2}  + (\frac{n\pi}{L p^2})^2  \delta_1  } \\
\pm(\frac{n\pi}{L p^2})  \delta_2 
\ee
where the finite mass gap is independent of $L$: 
$$
\Delta=\frac{\sqrt{2\pi g  \big(  k_1(k_1-k_2)v_1+2(k_3+p)(k_3 v_1-k_1 v_2) \big)}}  {|\gcd(k_1,k_3+p)|},
$$
$\delta_1=(k_1-k_2)^2v_1^2+4(k_3 v_1-k_1 v_2)(k_3 v_1-k_2 v_2)$ and $\delta_2= v_1(k_1+k_2)-v_2(2k_3)$.

To count the boundary GSD, for a generic $K_{2 \times 2}$ Abelian topological order on a disk, take $\ell_a^{} \in \Gamma^\partial_{}$ in Eq.\;(\ref{eq:l}) without losing generality (the same argument for $\Gamma^{\partial'}_{}$),\cite{note_1} we have 
$$
{P_{\phi_1}}=I_1\ell_{a,1}^{} +j_A  \frac{\ell_{a,1}^{}}{|\gcd(\ell_a^{} )|},\;\;
{P_{\phi_2}}=I_1\ell_{a,2}^{} +j_A  \frac{\ell_{a,2}^{}}{|\gcd(\ell_a^{} )|}.
$$
The total anyon charge for each branch needs to conserve, but a single boundary of a disk has no other boundaries to locate transported anyons. This implies: $j_A=0$, there is no different topological sector induced by transporting anyons
, thus $\GSD=1$.

On the other hand, if the topology 
is replaced by a cylinder with the top $\partial_1$ and the bottom $\partial_2$ boundaries shown in \figref{GSD_1or2orp}, 
when the gapping terms from {\it boundary gapping lattice} $ \Gamma^\partial_{}$ are chosen to be the same,
the Hilbert space on $P_{\phi}$ lattice is:
\bea
&&{P_{\phi_1}}=I_1 \ell_{a,1}^{} +j_A  \frac{\ell_{a,1}^{}}{|\gcd(\ell_{a} )|},\;
{P_{\phi_2}}=I_1 \ell_{a,2}^{} +j_A  \frac{\ell_{a,2}^{}}{|\gcd(\ell_{a} )|}, \;\;\;\;\;\;\;\;\label{eq:Pu}\\
&&{P_{\phi_3}}=I_2 \ell_{a,1}^{} +j_B  \frac{\ell_{a,1}^{}}{|\gcd(\ell_{a} )|},\;
{P_{\phi_4}}=I_2 \ell_{a,2}^{} +j_B  \frac{\ell_{a,2}^{}}{|\gcd(\ell_{a} )|}. \;\;\;\;\;\;\;\;\label{eq:Pd}
\eea
Anyon fusion rule and charge conservation for each branch constrains 
$( {P_{\phi_1}}+{P_{\phi_3}}, {P_{\phi_2}}+{P_{\phi_4}})$ belongs to the $\Gamma_e$ electron lattice: $( {P_{\phi_1}}+{P_{\phi_3}}, {P_{\phi_2}}+{P_{\phi_4}}) \in \Gamma_e$.
With $|\gcd({\ell_a}^{} )|=p$, it implies $j_A=-j_B (\bmod \;p)$. $0\leq j_A (\bmod \;p)<p$ has $p$ different topological sectors induced by different $j_A$. When $\Delta j_A /p \in \mathbb{Z}$, 
it transports non-fractionalized particles (e.g. electrons), so it brings back to the same topological sector. 
Count the number of distinct sectors, i.e. ground states, we find $\GSD=p$.

If gapping terms on two boundaries of a cylinder are chosen to be different: $\Gamma^\partial_{}$ for  $\partial_1$, $\Gamma^{\partial'}_{}$ for $\partial_2$, 
we revise the second line of Eq.\;(\ref{eq:Pd}) to
\be
{P_{\phi_3}}=I'_2 \ell'_{a,1} +j'_B  \frac{\ell'_{a,1}}{|\gcd(\ell'_{a} )|},\;
{P_{\phi_4}}=I'_2 \ell'_{a,2} +j'_B  \frac{\ell'_{a,2}}{|\gcd(\ell'_{a} )|}.
\ee
Anyon fusion rules and anyon conservation imply: $(\frac{k_1}{|\gcd(\ell_a^{} )|}j_A+\frac{k_3+p}{|\gcd(\ell'_{a} )|}j'_B, \frac{(k_3+p)}{|\gcd(\ell_a^{} )|}j_A+\frac{k_2}{|\gcd(\ell'_{a} )|} j'_B) \in \Gamma_e$.
This constraint gives a surprise. 
For example, $K_{Z_p}= \bigl( {\begin{smallmatrix}  
0 &p \\
p & 0 
 \end{smallmatrix} } \bigl)
$, we obtain $\GSD=1$. However, when $K_{\diag,p}=
\bigl( {\begin{smallmatrix}  
p &0 \\
0 & -p 
 \end{smallmatrix} } \bigl)$,
we obtain $\GSD=1$ for $p\in$ odd, but $\GSD=2$ for $p\in \even$.
This provides a new approach that one can distinguish two types of orders $K_{Z_p}$ and $K_{\diag,p}$ when $p$ is even by measuring their boundary GSD.

We illustrate this result in an intuitive way in \figref{GSD_1or2orp}. 
When boundary types and boundary gapping lattices $\Gamma^\partial$ 
are the same on two sides of the cylinder, \figref{GSD_1or2orp}(a) is enough to explain $\GSD=p$ , where fractionalized anyons transport
from the bottom to the top. For $K_{Z_p}$ case, say $\Gamma^{\partial_{1}}=\Gamma^{\partial_{2}}=\Gamma^\partial_{}=\{n(p,0)\}$, 
there are quasiparticle qp$_1$ with $\ell_a=j_A(1,0)$ for $0 \leq j_A \leq p-1$.
For $K_{\diag,p}$ case, say $\Gamma^{\partial_{1}}=\Gamma^{\partial_{2}}=\Gamma^\partial_{}=\{n(p,p)\}$, qp$_1$ with $\ell_a=j_A(1,1)$ for $0 \leq j_A \leq p-1$.
This accounts all $p$ sectors.

\begin{figure}[!h]
{\includegraphics[width=.40\textwidth]{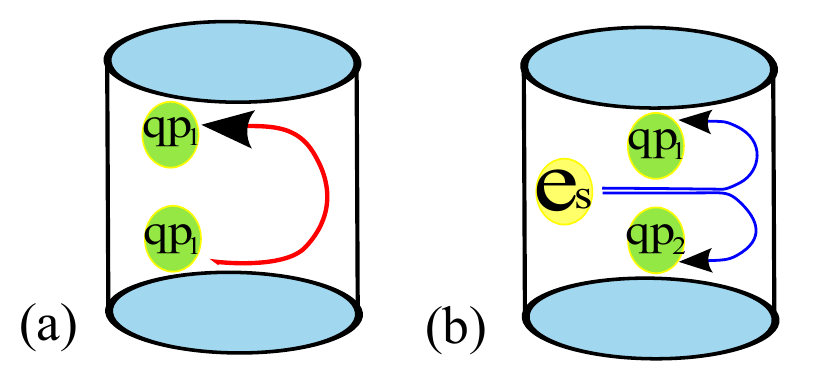}}
\caption{
(a) Anyon (qp$_1$) is transported from the bottom to the top of the cylinder.
(b) Physical non-fractionalized excitation e$_s$ splits into a pair of anyons (qp$_1$ to the bottom, qp$_2$ to the top).
}
\label{GSD_1or2orp}
\end{figure}

Let us first do a preliminary analysis, when \emph{boundary gapping lattices} $\Gamma^\partial$ are different. Naively, \figref{GSD_1or2orp}(a) is not allowed for the fractionalized anyon transport. 
\figref{GSD_1or2orp}(b) is crucial to account for the second ground state of $K_{\diag,p}$ at $p\in \even$. Let us take  
$\Gamma^{\partial_{1}}=\Gamma^\partial_{}=\{n(p,p)\}$ and $\Gamma^{\partial_{2}}=\Gamma^{\partial'}_{}=\{n(p,-p)\}$,
where e$_s$ represents a non-fractionalized particle $\ell=(p,0)$, while
qp$_1$ with $\ell_a=(p/2,p/2)$ and qp$_2$ with $\ell_a=(p/2,-p/2)$ at $p\in \even$ are allowed fractionalized anyons (with integer unit of anyon charge). 
This process switches the ground state to a different sector, so $\GSD=2$.
However, fractionalized anyons transport in \figref{GSD_1or2orp}(b) is not allowed for $K_{Z_p}$ with different boundary types on two sides of the cylinder, which results in $\GSD=1$. 

Next let us do a more careful analysis, taking into account the \emph{boundary gapping conditions} defined through the set of condensed anyons (namely, the third column of Table \ref{table:bGSD_TO} in Sec.\ref{sec:bGL_bGC}). 
In this case, we learn that the two boundary gapping lattices for $K_{\diag,2}$ give rise to the same boundary gapping condition:
$\{ (1,1),(2,2),\dots \}$. 
The anyon transport picture of \figref{GSD_1or2orp}(a) and (b) represent the same kind of transport, since
$\ell_a=(1 \pmod{2},1 \pmod{2})=(1 \pmod{2},-1 \pmod{2})$.
However, for $K_{\diag,4}$, the two boundary gapping lattices $\Gamma^{\partial_{1}}=\Gamma^\partial_{}=\{n(4,4)\}$
and $\Gamma^{\partial_{2}}=\Gamma^{\partial'}_{}=\{n(4,-4)\}$ on a cylinder represent different boundary gapping conditions:
$\{ (1,1),(2,2),(3,3),\dots \}$ and 
$\{ (1,3),(2,2),(3,1),\dots \}$ respectively.
The anyon transport picture of \figref{GSD_1or2orp}(a) and (b) for $K_{\diag,4}$ still represent the same kind of transport, since
 $\ell_a=(2 \pmod{4},2 \pmod{4})=(2 \pmod{4},-2 \pmod{4})$.
More generally, we have $\ell_a=(p/2 \pmod{p},p/2 \pmod{p})=(p/2 \pmod{p},-p/2 \pmod{p})$.
To summarize, for $K_{\diag,p}$ with $p$ is an even integer, 
the boundary gapping conditions $\{ (1,1),(2,2), \dots, (p/2,p/2), \dots \}$ and $\{ (1,p-1),(2,p-2), \dots, (p/2,p/2), \dots \}$
on two sides of a cylinder give rise to GSD=2. The second ground state is obtained through transporting the $(p/2,p/2)$ anyon from one side to the other side of the cylinder.

The result remarks that only the fusion algebra (both $K_{Z_p}$ and $K_{\diag,p}$ have the doubled fusion algebra $(\mathbb{Z}_p)^2$) is not sufficient enough to determine the boundary GSD.

\section{Surgery to glue cylinders to form a genus $g$ Riemann surface with punctures} \label{sec:genus_g_R}

Here we show how to glue the boundaries of punctured cylinders to form a genus $g$ Riemann surface with punctures, and determine its GSD of the topological order
on the surface.

For a genus $g$ Riemann surface with $\eta'$ punctures (\figref{hole}(b) and \figref{Genus_glue}(c)), we start from  
\figref{Genus_glue}(a), a number of $g$ cylinders drilled with a total puncture number $\eta=\eta'+2g+2(g-1)$, 
where $2g$ count two punctures on top and bottom for each cylinder (h$_{i,T}$ and h$_{i,B}$, $1\leq i \leq g$), drill an extra puncture on both the $1$st (h$_{1,L}$) and the last $g$th cylinder (h$_{g-1,R}$), and drill two extra punctures (h$_{j-1,R}$ and h$_{j,L}$) for the $j$-th cylinder for $2\leq j \leq g-1$. There are thus $2(g-1)$ extra punctures. Glue the boundaries of h$_{j,L}$ and h$_{j,R}$ together for $1\leq j \leq g-1$, and glue h$_{i,T}$ and h$_{i,B}$ together for $1\leq i \leq g$, results in \figref{Genus_glue}(c).

Use Eq.\;(\ref{gsd}) to account for the part with glued boundaries  ($1\leq \alpha \leq 2g+2(g-1)$) which contributes a factor of $ |\det K|^g$, 
and redefine particle hopping lattices $L_{qp \bigcap e}$ and $\bigoplus_{\alpha'} \Gamma^{\partial_{\alpha'}}_{} $ only for unglued boundaries ($1\leq \alpha' \leq \eta'$), we derive
\be
\GSD \leq |\det K|^g \cdot \; \vline \frac{ L_{qp \bigcap e}}{ \bigoplus^{\eta'}_{\alpha'=1} \Gamma^{\partial_{\alpha'}}_{}  } \vline \;. 
\ee
For a genus $g$ Riemann surface ($\eta'=0$), this becomes 
$
\GSD \leq |\det K|^g \nonumber
$, 
where $\rank(K)=N$ for a closed manifold case is relaxed to any natural number $\mathbb{N}$, since we still can create $2N$ modes by gluing two boundaries each with an odd number $N$ of edge modes.
This works for both odd and even number of branches.
The inequalities here are due to 
different choices of boundary gapping conditions for 
glued boundaries. 

\begin{widetext}
\begin{center}
\begin{figure}[!h]
{\includegraphics[width=.55\textwidth]{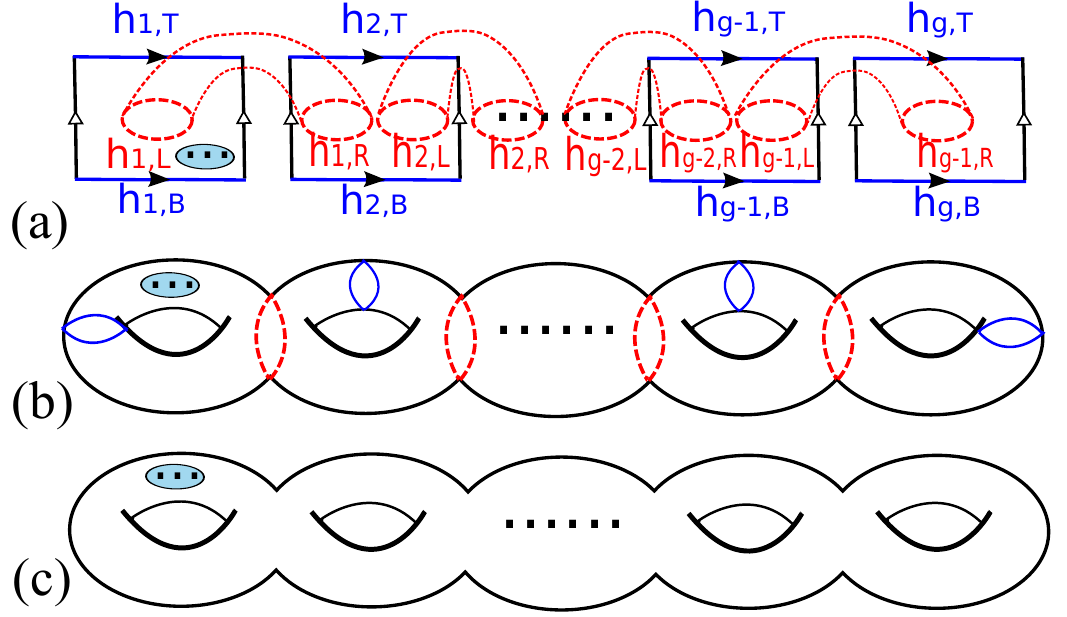}}
\caption{Glue the punctured cylinders to form a genus $g$ Riemann surface with $\eta'$ punctures. Start from (a), firstly identify left and right $\vartriangleright$ arrows of each square to form a 
number $g$ of punctured cylinders.
Then glue h$_{j,L}$ and h$_{j,R}$ (red dotted circles) together for $1\leq j \leq g-1$, and glue h$_{i,T}$ and h$_{i,B}$ (blue arrows) together for $1\leq i \leq g$, which yields (b), equivalently as a genus $g$ Riemann surface (c). The extra $\eta'$ punctures are indicated here as a shaded blue puncture in the left most handle.}
\label{Genus_glue}
\end{figure}
\end{center}
\end{widetext}

\section{Number of types of boundary gapping conditions} \label{sec:Ng}
For the number of types of boundary gapping conditions $\mathcal{N}^\partial_g$ for a $K$-matrix Chern Simons theory,
we have discussed the subtle difference between the boundary gapping lattices and the boundary gapping conditions in Sec.\ref{sec:bGL_bGC}.
For $\rank(K)=2$, we showed that there are two types of boundary gapping lattices, 
but the $\mathcal{N}^\partial_g$ in Table \ref{table:bGSD_TO} can be $1,2,3,$ etc.
For $\rank(K) \geq 4$, even though there can be infinite types of boundary gapping lattices, but the $\mathcal{N}^\partial_g$ can be {\it finite}. 

Let us consider a specific case with $|\det K|=1$, where the canonical form of this unimodular indefinite symmetric integral $K_{N\times N}$ matrix exists.\cite{canonical}
For the odd matrix (where the quadratic form has some odd integer coefficient, so the system is fermionic with fermionic statistics),
the canonical form is composed by $N/2$ blocks of
\be \label{eq:K_f}
\bigl( {\begin{smallmatrix} 
1 &0 \\
0 & -1 
\end{smallmatrix}}  \bigl) \oplus \bigl( {\begin{smallmatrix} 
1 &0 \\
0 & -1 
\end{smallmatrix}}  \bigl) \oplus \dots
\ee 
along the diagonal blocks of $K_{N\times N}$.
For the even matrix (where the quadratic form has only even integer coefficients, so the system is bosonic with only bosonic statistics),
the canonical form is composed by blocks of
\be  \label{eq:K_b}
\bigl( {\begin{smallmatrix} 
0 &1 \\
1 & 0  
\end{smallmatrix}} \bigl) \oplus \bigl( {\begin{smallmatrix} 
0 &1 \\
1 & 0  
\end{smallmatrix}} \bigl) \oplus \dots
\ee 
and a set of all positive (or negative) coefficients $E_8$ lattices, $K_{E_8}$, 
\be
  K_{E_8}=  \begin{pmatrix}
      2 & -1 & 0 & 0 & 0 & 0 & 0 & 0\\
      -1 & 2 & -1 & 0 & 0 & 0 & 0 & 0 \\
      0 & -1 & 2 & -1 & 0 & 0 & 0 & -1 \\
      0 & 0 & -1 & 2 & -1 & 0 & 0 & 0 \\
      0 & 0 & 0 & -1 & 2 & -1 & 0 & 0 \\
      0 & 0 & 0 & 0 & -1 & 2 & -1 & 0 \\
      0 & 0 & 0 & 0 & 0 & -1 & 2 & 0 \\
      0 & 0 & -1 & 0 & 0 & 0 & 0 & 2 \\
    \end{pmatrix}
\ee
along the diagonal blocks of $K_{N\times N}$.
A positive definite $K_{E_8}$ with eight chiral bosons cannot be gapped out. 
Thus, in order to have non-chiral states,  
the even matrix canonical form must be composed by $N/2$ blocks of
$\bigl( {\begin{smallmatrix} 
0 &1 \\
1 & 0 
\end{smallmatrix}}\bigl)$.

Now let us revisit the number of boundary gapping conditions $\mathcal{N}^\partial_g$ in this canonical form when $|\det K|=1$. 
We had claimed under the boundary fully gapping rules ({\bf1})({\bf2})({\bf3})({\bf4}) when $\rank(K)\geq 4$, the number of types of boundary gapping lattices can be infinite. 
Now for the boundary gapping conditions, we should
identify the boundary gapping lattices via the particle lattice by Eq.(\ref{eq:identify gapping lattices}) of Sec.\ref{sec:bGL_bGC}.
This identification modifies our result for the fermionic system of the $K$ matrix in Eq.(\ref{eq:K_f}) 
and for the bosonic system of the $K$ matrix in Eq.(\ref{eq:K_b}) to 
$$\mathcal{N}^\partial_g=1.$$ 
Due to the integer lattice identification of Eq.(\ref{eq:identify gapping lattices}),
we cannot distinguish $0$ and $1$ due to the module $1$ identification by the column vector of $K$ matrices in Eq.(\ref{eq:K_f}) and Eq.(\ref{eq:K_b}). 
All trivial particles, bosons for bosonic system or fermions for fermionic systems of $|\det K|=1$, are identified.
Without symmetry or symmetry-breaking, all the boundary types with the condensation of trivial particles are thus identified.




\begin{thebibliography}{99}
\bibitem{lattice-C-S}
Later we will show that for the Abelian topological orders, 
$\Gamma^{\partial_\alpha}$ and $\Gamma^{\partial_\alpha}_{qp}$ are the charge lattice Hilbert space of the quantized Chern-Simons theory.
\bibitem{direct_sum}
For Abelian group, finite direct product is finite direct sum, so $\bigoplus_\alpha \Gamma^{\partial_\alpha}_{} =\prod_\alpha \Gamma^{\partial_\alpha}_{} $, with $\alpha$ a finite number of punctures.
Direct sum is indeed the coproduct in the category of abelian groups.
\bibitem{braiding}
Peculiarly simply apply Eq.\;(\ref{gsd}) to surfaces without genus (\figref{hole}(a)) and with genus (\figref{hole}(b)) cases gives the same GSD, if both have the same boundary number and boundary conditions.
However by gluing techniques, we learn from Eq.\;(\ref{gsdG}) there can be a factor contribution($|\det K|^g$) from the genus topology.
So the nontrivial winding in genus may modify Eq.\;(\ref{gsd}), where braiding rules may involve.
\end{thebibliography}
\begin{thebibliography}{20}
\bibitem{direct_sum}
For Abelian group, a finite direct product is a finite direct sum, so $\bigoplus_\alpha \Gamma^{\partial_\alpha}_{} =\prod_\alpha \Gamma^{\partial_\alpha}_{} $, 
with $\alpha$ as the finite number of boundaries. 
The direct sum is indeed the coproduct in the category of abelian groups.
\bibitem{braiding}
Peculiarly simply apply Eq.\;(\ref{gsd}) to surfaces without genus (\figref{hole}(a)) and with genus (\figref{hole}(b)) cases gives the same GSD, if both have the same boundary number and boundary conditions.
However by gluing techniques, we learn from Eq.\;(\ref{gsdG}) there can be a factor contribution($|\det K|^g$) from the genus topology.
So the nontrivial winding in genus may modify Eq.\;(\ref{gsd}), where braiding rules may get involved.
On the other hand, braiding rules are implicitly involved when we define gapped boundaries through the particle condensation with the trivial braiding statistics among
the condensed particle and condensed quasiparticles.
\bibitem{note_1}
Notice if we take $|n|\geq 2$ or $|n'|\geq2$ larger-size hopping term for $n {\ell_{a,I}^{}} \in \Gamma^\partial_{}$ ,$n' {\ell'_{a,I}}\in \Gamma^{\partial'}_{}$, the lattice can break into more sublattices of different sectors: the calculation on a disk is altered to $\GSD=|n|$. However, generically
no symmetry forbids the $n=n'=1$ case (the least-size hopping terms), whose dominant cosine potential can minimize the energy. 





\begincomment
\end{thebibliography}
\end{document}